\documentclass
[aps,prc,twocolumn,showpacs,amsmath,floatfix,superscriptaddress]{revtex4}
\usepackage{color,graphicx,ulem,subfig}
\usepackage{mathptmx}                
\usepackage{dcolumn}                 
\usepackage{bm}                      
\usepackage{epsfig,amsmath}
\usepackage{xcolor}

\begin{document}
\title{Modification of magicity towards the dripline and its impact on 
electron-capture rates for stellar core-collapse}

\author{Ad. R. Raduta}
\affiliation{IFIN-HH, Bucharest-Magurele, POB-MG6, Romania}
\author{F. Gulminelli}
\affiliation{CNRS/ENSICAEN/LPC/Universit\'e de Caen Basse Normandy, 
UMR6534, F-14050 Caen c\'edex, France}
\author{M. Oertel}
\affiliation{LUTH, CNRS, Observatoire de Paris, Universit\'e Paris Diderot, 
5 place Jules Janssen, 92195 Meudon, France}

\begin{abstract}
The importance of microphysical inputs from laboratory nuclear
experiments and theoretical nuclear structure calculations in the
understanding of the core collapse dynamics, and the subsequent
supernova explosion, is largely recognized in the recent literature.
In this work, we analyze the impact of the masses of very neutron
rich nuclei on the matter composition during collapse, and the
corresponding electron capture rate.  To this aim, we introduce an
empirical modification of the popular Duflo-Zuker mass model to
account for possible shell quenching far from stability, and study the
effect of the quenching on the average electron capture rate.  We show
that the preeminence of the $N=50$ and $N=82$ closed shells in the
collapse dynamics is considerably decreased if the shell gaps are
reduced in the region of $^{78}$Ni and beyond. 
As a consequence, local modifications of the overall electron capture rate up to 
30\% can be expected, with integrated values strongly dependent on the 
stiffness of magicity quenching and progenitor mass and potential important consequences 
on the entropy generation, the neutrino emissivity, and the mass of the core
at bounce.
Our work underlines the importance of new experimental
measurements in this region of the nuclear chart, the most crucial
information being the nuclear mass and the Gamow-Teller strength.
Reliable microscopic calculations of the associated elementary rate,
in a wide range of temperatures and electron densities, optimized on
these new empirical information, will be additionally needed to get
quantitative predictions of the collapse dynamics.
\end{abstract}


\today

\maketitle

\section{Introduction}
 It is nowadays well recognized that reliable nuclear physics inputs
 are essential for realistic simulations of many astrophysical
 phenomena. Depending on the particular time scales and the resulting
 equilibrium conditions, this can be an equation of state (EoS) and
 individual (nuclear) reactions, respectively. Due to the relatively
 long timescale of reactions mediated by the weak interaction, the
 latter play an important role in many sites, for instance the late
 stages of massive star evolution \cite{Aufderheide94, Heger01,
   Martinez14}, thermonuclear \cite{Iwamoto99, Brachwitz00} and
 core-collapse supernovae \cite{Hix03, Langanke03, Janka07},
 nucleosynthesis and energy generation in X-ray bursts and other
 rp-process sites \cite{Schatz98}, the accreting neutron star
 crust~\cite{Gupta06,Schatz14}, and neutron star
 mergers~\cite{Mendoza14, Goriely15}.

Within this paper, we will concentrate on core-collapse supernovae
(CCSNe).  Except for very low-density matter encountered in the outer
layers, time scales are such that strong and electromagnetic
interactions are in equilibrium. Therefore, at any time, the
composition of matter can be calculated as a function of the local
temperature ($T$), baryon number density ($n_B$) and proton fraction
($Y_p$) assuming Nuclear Statistical Equilibrium (NSE) (see
e.g.~\cite{Janka07}). On the contrary, weak interactions can in
general not be considered in equilibrium and individual reactions
rates are crucial to determine the local proton fraction.  In
particular, electron capture determines $Y_p$ in the first stages of
the collapse, the associated size of the homologous core, and has thus an
impact on the consequent explosion dynamics
\cite{Bethe79,Bruenn85,Hix03,Janka07,
Hempel12,Langanke14,Sullivan15}.

For these simulations, the time and space dependent electron capture
rates are obtained for a given $(n_B, T, Y_p)$ by folding the NSE
nuclear distribution with the capture rates on individual nuclei. The
microphysics uncertainties on the rates thus originate both from the
uncertainties on the NSE distribution, and on those associated with
the individual rates.

 A very complete study on the sensitivity of core collapse dynamics to
 variations of electron capture rates on medium-heavy nuclei has
 recently been performed~\cite{Sullivan15}. The authors thereby
 concentrate on the latter aspect, modifying the rates on individual
 nuclei according to present uncertainties, using a comprehensive set
 of progenitors and EoS. It was shown that important variations in the
 mass of the inner core at bounce and in the peak neutrino luminosity
 have to be expected. The variations induced by the modified electron
 capture rates are five times larger than those induced by the
 uncertainty on the progenitor model, showing the importance of an
 increased reliability of nuclear physics inputs. In this same work it
 was clearly shown that the results are most sensitive to the region
 around the $N=50$ shell closure for very neutron rich nuclei ($74\leq
 A \leq 84$).

In this work we address the complementary aspect of uncertainties
associated with the nuclear distribution.  Specifically, nuclear
magicity, as incorporated in mass models within currently available
EoS, is known to be deeply modified in very neutron  rich nuclei
~\cite{Porquet08} and the  pronounced shell closures at 
$N=50$ and $N = 82$ are expected to
 be quenched far from stability.
We study here the impact of such a possible quenching on the matter
composition during collapse, and the associated modification of the
electron capture rate.  We show that modifications of the
electron capture rate of up to 30\% are possible in the case
of strong shell quenching. 

Unlike~Ref.~\cite{Sullivan15}, in this exploratory calculation we do
not aim at computing the complete time evolution of $Y_p$, nor the
associated modification of the inner core mass and the neutrino
luminosity. Such a complete simulation would require a consistent
model for the mass and electron capture rates as determined by Fermi
and Gamow-Teller transition strengths for the relevant nuclei. This
is beyond the scope of the present paper. However, from general arguments
and the very detailed results of Ref.~\cite{Sullivan15}, we expect
that an increased capture rate leads to enhanced neutrino cooling, an
accelerated collapse, and a higher inner core mass at bounce.  The
main message of the present work is to stress the need of structure
information and mass measurements, particularly around the $^{78}$Ni
region. From the theory viewpoint, it appears very important to have
microscopic and consistent calculations of both masses and weak rates.

The paper is organized as follows. In Section \ref{sec:sec2} the
different microphysical inputs entering the {electron capture} rate
calculations are presented, namely the individual rates and the
equation of state.  We show that for the latter, the mass model
represents the key ingredient.  The  {influence of the mass model on
} the capture rates during core collapse is
discussed in Section~\ref{sec:sec3}.  {To that end, we consider two
  representative collapse trajectories with typical thermodynamic
  conditions. }  An effective parametrization
simulating the possible shell quenching far from stability of the two
relevant shell closures $N=50$ and $N=82$ is introduced, and the
associated modification of the global rate is discussed. Section
\ref{sec:sec4} contains summary and conclusions.  {Throughout the
  paper we will use units such that $ k_B = 1$.}

\section{Ingredients of the rate calculation} \label{sec:sec2}

As mentioned above, for given thermodynamic condition, i.e. given
values of $(n_B, T, Y_p)$, the total electron capture rate is obtained
by folding individual rates with the nuclear distribution.  The latter
is obtained by NSE calculations which depend in turn on a number of
inputs and in particular on the masses of different nuclei, thus on
the mass model employed.  In this section we detail the different
ingredients employed in this work. We will start with the individual
rates in Sec~\ref{sec:individual}. The NSE model will be discussed in
Sec.~\ref{sec:NSE}. Sec.~\ref{sec:mass} will be devoted to
the mass model and special attention will be paid to the possible quenching of 
magicity in very neutron rich nuclei.

\subsection{Individual rates}
\label{sec:individual}

Concerning the individual rates, tabulated values are available from
large scale shell model calculations in the $sd$-shell \cite{Oda}
(OHMTS), and $fp$-shell~\cite{Langanke00,LMP} (LMP).  Since those
calculations are still nowadays numerically very demanding, for
heavier nuclei, the shell model rates are complemented with shell
model Monte Carlo and RPA calculations~\cite{Nabi} or an empirical
approach~\cite{Pruet} (PF).  In Ref.~\cite{Juodagalvis_2010} shell
model rates on additional nuclei have been calculated, however, they
are unfortunately not available as individual rates. These extended
calculations still comprise a limited range of elements, masses, temperature and
electron densities ($n_e = n_p$) and are not sufficient to cover
completely the typical conditions in the most central part of the
core collapse, nor the typical nuclei encountered in those situations.

We have therefore decided to use here the analytical equation
proposed by Langanke et al. (L03) in Ref. \cite{Langanke03}. 
It is based on a parametrization of the electron-capture rate as
a function of the ground state to ground state $Q$-value and
writes,
\begin{eqnarray}
\lambda_{EC}=\frac{\ln 2 \cdot B}{K} \left(\frac{T}{m_e c^2} \right)^5 
\left[ F_4(\eta)-2 \chi F_3(\eta)+\chi^2 F_2(\eta)
\right],
\label{eq:L03}
\end{eqnarray}
with $K=6146$ s, $\chi=(Q-\Delta E)/T$, $\eta=\chi+\mu_e/T$.
$T$ represents the temperature,
$m_e$ and  $\mu_e$  stand for electron rest mass and, respectively,
chemical potential. $F_i(\eta)$ denotes the relativistic Fermi integral,
$F_i(\eta)=\int_0^{\infty} dx x^k/(1+\exp(x-\eta))$.
$B$ stands for a typical (Gamow-Teller plus forbidden) matrix element and
$\Delta E=E_f-E_i$ represents the energy difference between the parent state $E_i$
and the daughter state $E_f$.
Their values, obtained by fitting microscopic data, are $B=4.6$ MeV and $\Delta E=2.5$ MeV. 
For electron capture rates on protons we use the results of Ref.~\cite{FFN82}.

The first approximation of weak interaction rates in terms lepton chemical potential 
and $Q$-value was proposed by Fuller, Fowler and Newman in the pioneering Ref. \cite{FFN85}
in the form $\lambda_{EC}=\ln 2 I_e/\langle ft \rangle_e$, where $I_e$ is the space factor.
Nuclear structure effects were there approximately accounted for via the 
effective log(ft)-value. For EC-rates three values have been proposed based on the
neutron and proton numbers of the parent nucleus (upon which nuclei are classified in 
"blocked" ($N \geq 40$ or $Z \leq 20$) and "unblocked" ($N<40$ and $Z>20$))
and the relation between electron chemical potential and $Q$-value. For "unblocked" nuclei EC was
supposed to occur mainly through the Gamow-Teller resonances which are roughly at an excitation
of 3 MeV above the daughter ground state. 

\begin{figure}
\begin{center}
\includegraphics[angle=0, width=0.99\columnwidth]{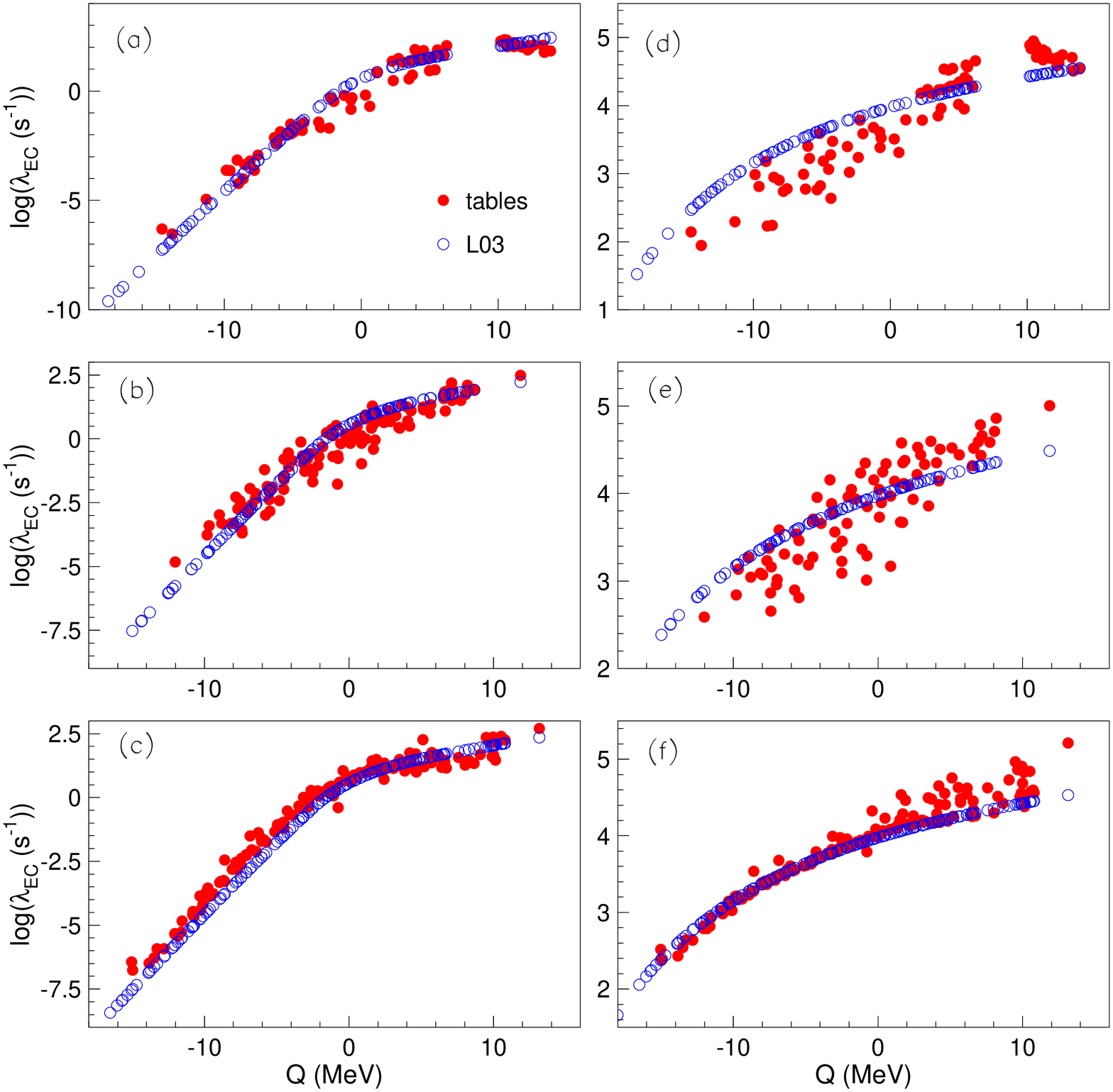}
\end{center}
\caption{(Color online) EC rates: comparison between L03 (open blue circles) and 
OHMTS~\cite{Oda}  ($17 \leq A \leq 39$) (a,d),
LMP~\cite{Langanke00,LMP} ($45 \leq A \leq 65$) (b,e) 
and PF \cite{Pruet} ($65 \leq A \leq 80$) (c,f) tables
(solid red circles) for $T$=0.68 MeV, $n_B=1.32 \cdot 10^{-6}$ fm$^{-3}$, $Y_e$=0.447 (a,b,c)
and $T$=1.30 MeV, $n_B=1.12 \cdot 10^{-4}$ fm$^{-3}$, $Y_e$=0.361 (d,e,f).
} 
\label{fig:EC_table_L03}
\end{figure}

The validity of the analytic equation L03 has been recurrently addressed in the literature
(see e.g. \cite{Langanke03,Sullivan15}) and proved to offer a fair approximation under
thermodynamical conditions relevant for core-collapse.
Scatter of microscopic rates around the $Q$-dependence at low values of temperature and
electron density were interpreted in Ref. \cite{Langanke03} as indicating that several
parent and daughter states with different transition strengths contribute to the process.
On the contrary, better agreement at higher values of temperature and electron density
was attributed to a  value of the electron chemical potential high enough to make
EC-rate independent of the strength distribution.
Fig. \ref{fig:EC_table_L03} addresses the same issue by confronting the predictions 
of eq. (\ref{eq:L03}),
plotted as a function of reaction heat $Q$, with those obtained by
linear interpolation of more microscopic and sophisticated weak interaction rate tables
OHMTS~\cite{Oda} (panels (a) and (d)), 
LMP~\cite{Langanke00,LMP} (panels (b) and (e)) 
and PF~\cite{Pruet} (panels (c) and (f)).
Two representative thermodynamic conditions taken from
the core-collapse trajectory of a $25M_{\odot}$ star in
Ref.~\cite{Juodagalvis_2010} are considered (see figure caption). 
As already evidenced by other authors, Fig. \ref{fig:EC_table_L03} shows 
a correct overall behavior of $\lambda_{EC}(Q)$ and, in what concerns the individual rates, 
important deviations. Quite remarkably, eq.(\ref{eq:L03}) offers a fair approximation
also outside the mass range for which it was designed.

Whenever exist the deviations underline the well known fact that,
in order to accurately describe astrophysical processes,
it will be very important to have fully microscopic calculations
covering the whole mass, charge, temperature and electron fraction domain. It
was recently demonstrated in Ref.~\cite{Sullivan15} that a global
arbitrary re-scaling of a factor 2,3 or 10 of the unknown rates leads
to important modifications in the dynamics of the collapse. Given the
purpose of this paper, which is to analyze the influence of the mass
extrapolations in the still experimentally unknown region of very
neutron rich nuclei, we have not tried to play with these rates, and
we consider that for our purpose the L03 equations are sufficiently accurate 
to provide a reference calculation.

\subsection {Extended NSE model}
\label{sec:NSE}

The model is based on a statistical distribution of compressible
nuclear clusters composed by $A$ nucleons ($N$ neutrons and $Z$
protons) immersed in a homogeneous background of self-interacting
nucleons and electrons.  The details of the model are explained
elsewhere \cite{Gulminelli15, Burrello15}; here we only recall the
main physical ingredients which are important for the present study.

The different thermodynamic quantities in the baryonic sector are
decomposed into the sum of a term pertaining to the nucleon gas, and a
term given by the contributions of the distribution of nuclei.  {Let
  us start with the gas contribution. In absence of clusters, it would
  simply be given by the} free energy density of homogeneous nuclear
matter at density $n_g=n_{gn}+n_{gp}$, asymmetry $\delta_g=(n_{gn}-n_{gp})/n_g$ and temperature $T$.
In the non-relativistic mean field approximation it reads ($q=n,p$):
\begin{eqnarray}
f_{HM}(n_g,\delta_g)&=& 
\sum_q g_q \int_0^\infty \frac{dp p^2}{2\pi^2 \hbar^3}  n_q^T \frac{p^2}{2m^*_q} + \nonumber \\ &+&\mathcal{E}_{pot}
- T s_{HM},
\label{eq:f_HM}
\end{eqnarray}
with the entropy density given by :
\begin{eqnarray}
s_{HM}(n_g,\delta_g)&=&-\sum_q g_q\int_0^\infty \frac{dp}{2\pi^2\hbar^3} p^2    [ n_q^T \ln n_q^T + \nonumber \\
&+& \left ( 1 - n_q^T \right ) \ln \left ( 1 - n_q^T \right ) ]
\label{eq:s_HM}
\end{eqnarray}
 In these equations,
$g_q=2$ is the spin degeneracy in spin-saturated matter, 
$n_q^T$ is the finite temperature occupation number at effective chemical 
potential $\tilde{\mu}_q=\mu_q-\partial \mathcal{E}_{pot}/
\partial n_{gq}$, 
$n_q^T=\left[1+\exp \left((p^2/2m^*_q - \tilde{\mu}_q)/T\right)\right]^{-1}$.

The choice for the energy density $\mathcal{E}_{pot}$ and the
effective nucleon mass $m^*_q$, containing the interaction effects,
defines the equation of state. For the applications shown in this
paper we have considered a large set of different well-known 
Skyrme functionals for these two quantities~\cite{Dan09}, 
which have been successfully confronted
with different nuclear structure data and are also
compatible with the most recent experimental constraints
on nuclear matter properties~\cite{Tsang12, Lattimer12, Lattimer14}. 

In the present NSE model, the free energy density of the nucleon gas 
is reduced with respect to the homogeneous gas expression, 
Eq.~(\ref{eq:f_HM}), in
order to account for the in-medium modification due to the finite
volume occupied by the clusters~\cite{Raduta10,Hempel10}. 
The result is:
\begin{equation}
f_{g}=f_{HM}  \left ( 1- 
 \sum_{N,Z} n^{(N,Z)}  \frac{A}{n_0(\delta)} \right ) ,  \label{eq:fgas}
\end{equation}
where the sum runs over the different clusters weighted by their
multiplicity per unit volume $n^{(N,Z)}$. $n_0(\delta)$ denotes the
saturation density of asymmetric matter evaluated at the cluster
asymmetry $\delta$, and accounts for the compressible character of the
clusters. The asymmetry $\delta$ differs from the cluster global asymmetry
$(N-Z)/A$ because of Coulomb and skin effects, and is additionally
modified to account for the influence of the external
gas~\cite{Papakonstantinou13,Gulminelli15}. 

We have checked that, within these phenomenological limits, the choice
of the interaction does not produce any sensible effect neither on the
composition of matter during collapse nor on the average capture rates.  
This is due to the fact that, during the collapse trajectories we have considered,
the density of the nucleonic gas and its neutron enrichment are not important enough
to make unbound nucleon energetics play a significant role.
This result is perfectly compatible with previous findings~\cite{Fischer14,Sullivan15}.

The cluster multiplicities are given by the self-consistent NSE expression:
\begin{eqnarray}
\ln n^{(N,Z)}  
&=&
-\frac{1}{T}\left (
F_T
-\mu_B A_e - \mu_p  Z_e
\right ), \label{mult_nse}
\end{eqnarray}
where we have defined the bound fraction of clusters by 
$A_e=A\left (1-n_g/n_0(\delta)\right )$, $Z_e=Z\left (1-n_{pg}/n_{0p}(\delta)\right )$,
with $n_{0p}=n_0(\delta)(1-\delta)/2$.
The chemical potentials can be expressed  
as a function of the gas densities only:
\begin{eqnarray}
\mu_B &\equiv& \frac{\partial f_{HM}}{\partial n_g};  \label{chem1}\\
\mu_p &\equiv& \frac{\partial f_{HM}}{\partial n_{gp}} \label{chem2}.
\end{eqnarray}

In eq.(\ref{mult_nse})  $F_T$ is the free energy of the cluster immersed in the nucleon gas:
\begin{eqnarray}
F_T(N,Z,n_g,\delta_g)&=&
-B - T \ln  \left (A_e^{\frac 3 2} c_T V_{tot} \right ) \nonumber \\
&-& f _{HM}(n_g,\delta_g) 
\frac{A}{n_0(\delta)}  \nonumber \\
&+&\delta F_{Coulomb}+\delta F_{surf},
\label{fenergy_cl_ws}
\end{eqnarray}
where the total volume $V_{tot}$ has been introduced.  Again, because of
the low gas densities considered in this application, the third and
fifth term in eq. (\ref{fenergy_cl_ws}), representing the in-medium
bulk and surface modification of the cluster free energy due to the
presence of the gas, turn out to be negligible. 

The most important
ingredient of the NSE model is thus given by the vacuum binding energy of the
cluster $B$, corrected by the well known electron screening
effect $\delta F_{coul}$, and, to less extent, the cluster
level density entering the temperature dependent degeneracy factor
$c_T$ given by:
\begin{equation}
c_T=\left (\frac{mT}{2\pi\hbar^2} \right)^{3/2} \int _0^{\langle S\rangle}dE \left [ D_{N,Z}(E)
\exp \left(- E/T \right) \right ],  \label{deg_bucurescu}
\end{equation}
where $D_{N,Z}$ is the density of states of the cluster, $\langle
S\rangle =\min(\langle S_n\rangle,\langle S_p\rangle)$ is the average
particle separation energy, $m$ is the nucleon mass. Concerning the
density of states,  we use a back-shifted Fermi gas model with
parameters fitted from experimental data~\cite{Egidy05}. The key
quantity of the model is thus the cluster binding energy, which we
shall  discuss in more detail in the following subsection.

\subsection {Mass model}
\label{sec:mass}

Consistency with the equation of state of the free nucleons in
principle demands that the cluster mass should be evaluated with the
same energy functional employed for the gas. 
This is indeed done in the most recent NSE models 
\cite{Raduta10,Hempel10,Blinnikov11,Typel10,Furusawa11}.  
However, no functional model sufficiently precisely reproduces nuclear ground states, 
and when experimentally measured nuclear masses are available, they are
preferentially used in all recent NSE
models~\cite{Gulminelli15,Burrello15,Hempel10,Furusawa11}. 
It is also important to stress
at this point that, independent of the nuclear matter parameters of
the EoS (symmetric nuclear matter incompressibility $K_\infty$, 
symmetry energy per nucleon at the saturation density of symmetric matter $J_0$, 
slope $L$ and curvature $K_{sym}$ of symmetry energy at symmetric matter saturation density), 
no functional model gives
completely reliable extrapolations of nuclear masses in the
neutron-rich region where experimental masses are not available. This
leads to a model dependence of the results, which is not related to
any unknown nuclear matter parameter, but rather to the poor
performance of the used Hartree-Fock or Thomas-Fermi approximations
in the evaluation of nuclear masses, especially in the very neutron
rich region.
Very few nuclear functional models exist, with parameters fitted with the same
degree of accuracy on infinite nuclear matter and to properties of
finite nuclei \cite{Geng05, Goriely02,Chamel11,Pearson13,Potekhin13}, 
and even in this
case the behavior of nuclear mass towards the dripline is subject to
great uncertainties.

Anyway, since in the collapse applications the energetics of the gas
plays a negligible role, the argument of consistency is not very
important. For this reason we have chosen to employ, whenever
experimental measurements are missing, the Duflo-Zuker mass
model~\cite{DZ10} which gives, within a microscopically inspired
formalism, the present best reproduction of measured nuclear masses.
For the unbound nucleon component the SLY4 \cite{SLY4} Skyrme effective 
interaction is used.
  
\begin{figure}
\begin{center}
\includegraphics[angle=0, width=0.35\columnwidth]{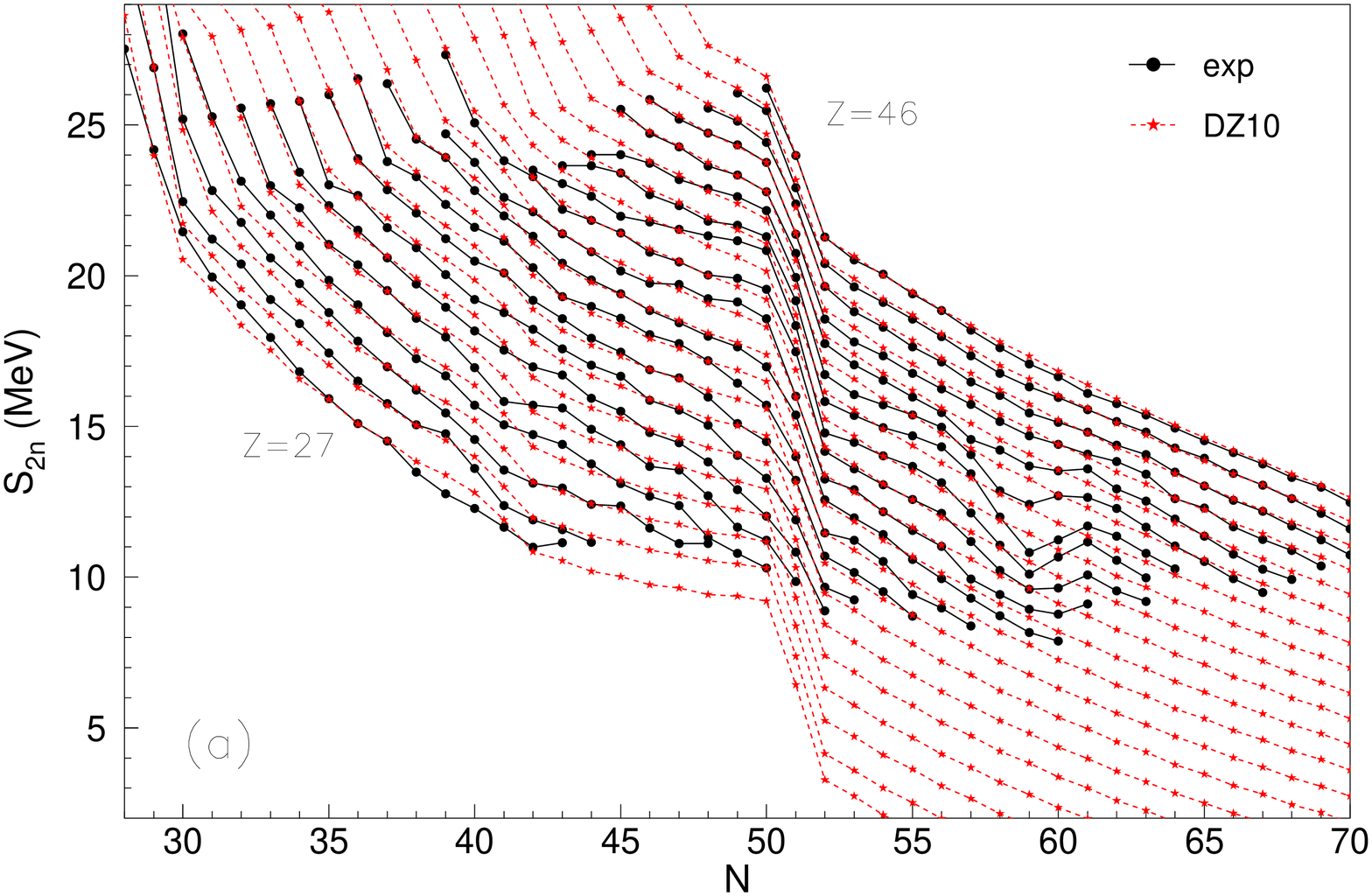}
\includegraphics[angle=0, width=0.35\columnwidth]{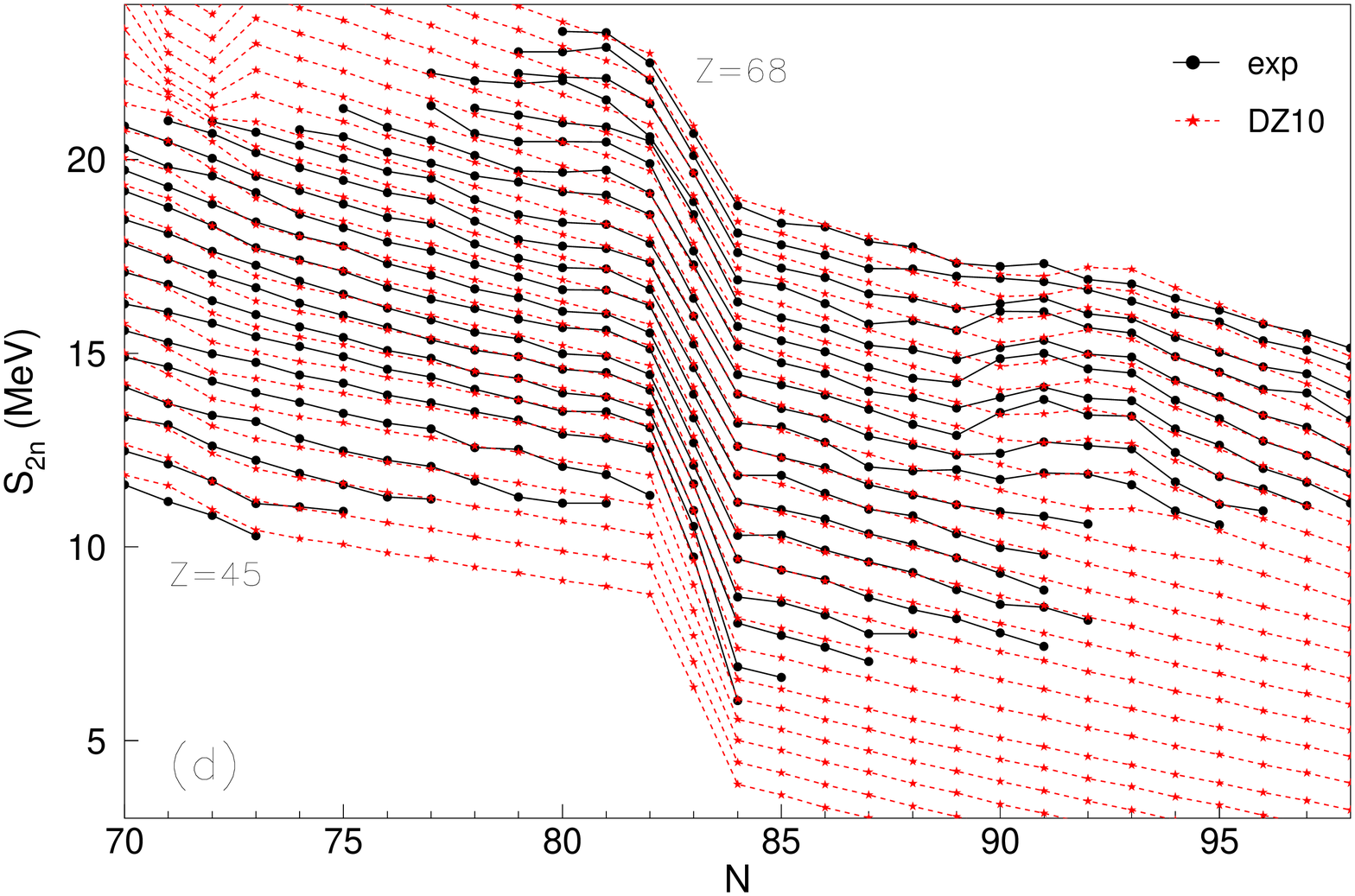}
\includegraphics[angle=0, width=0.35\columnwidth]{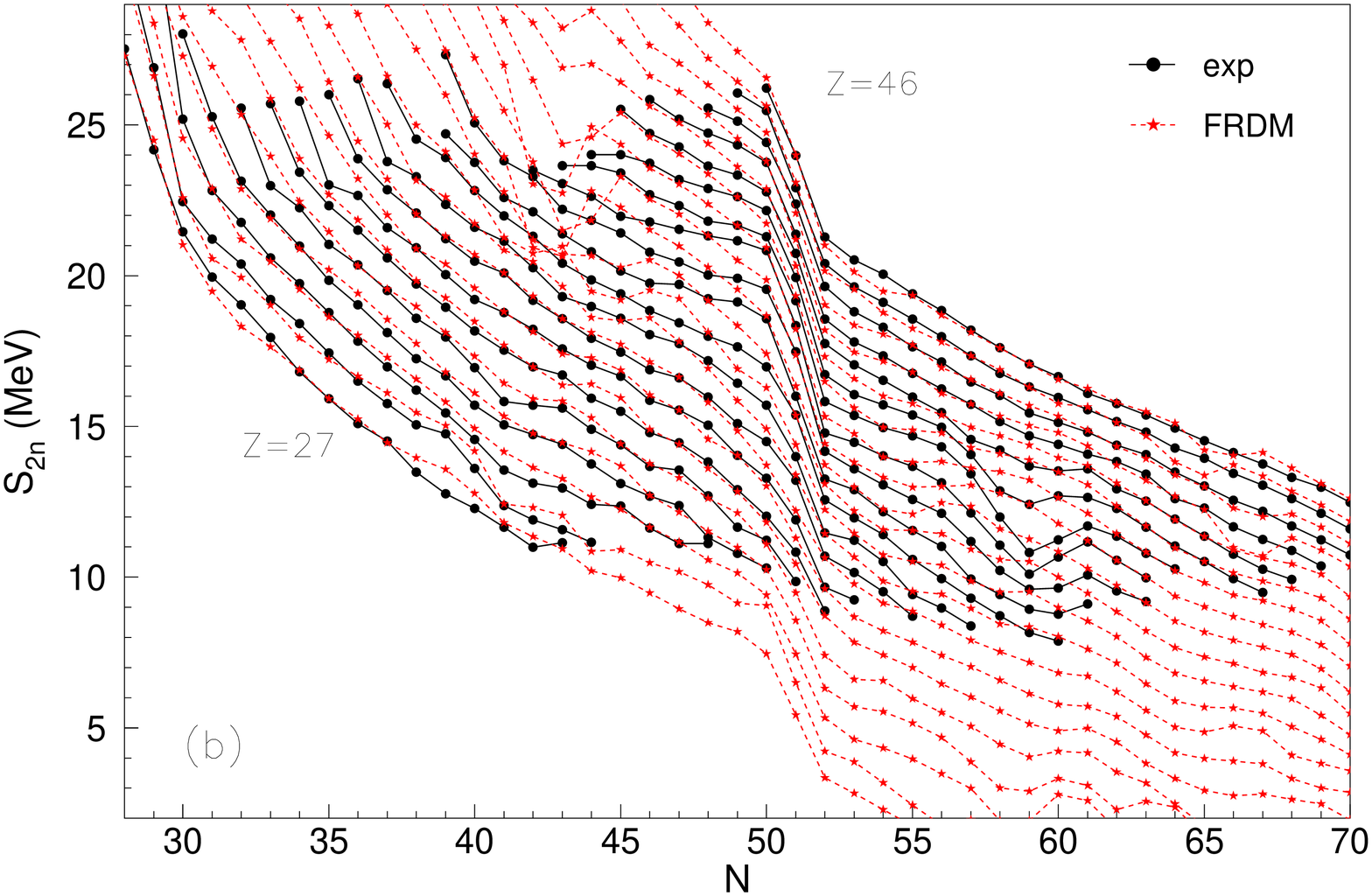}
\includegraphics[angle=0, width=0.35\columnwidth]{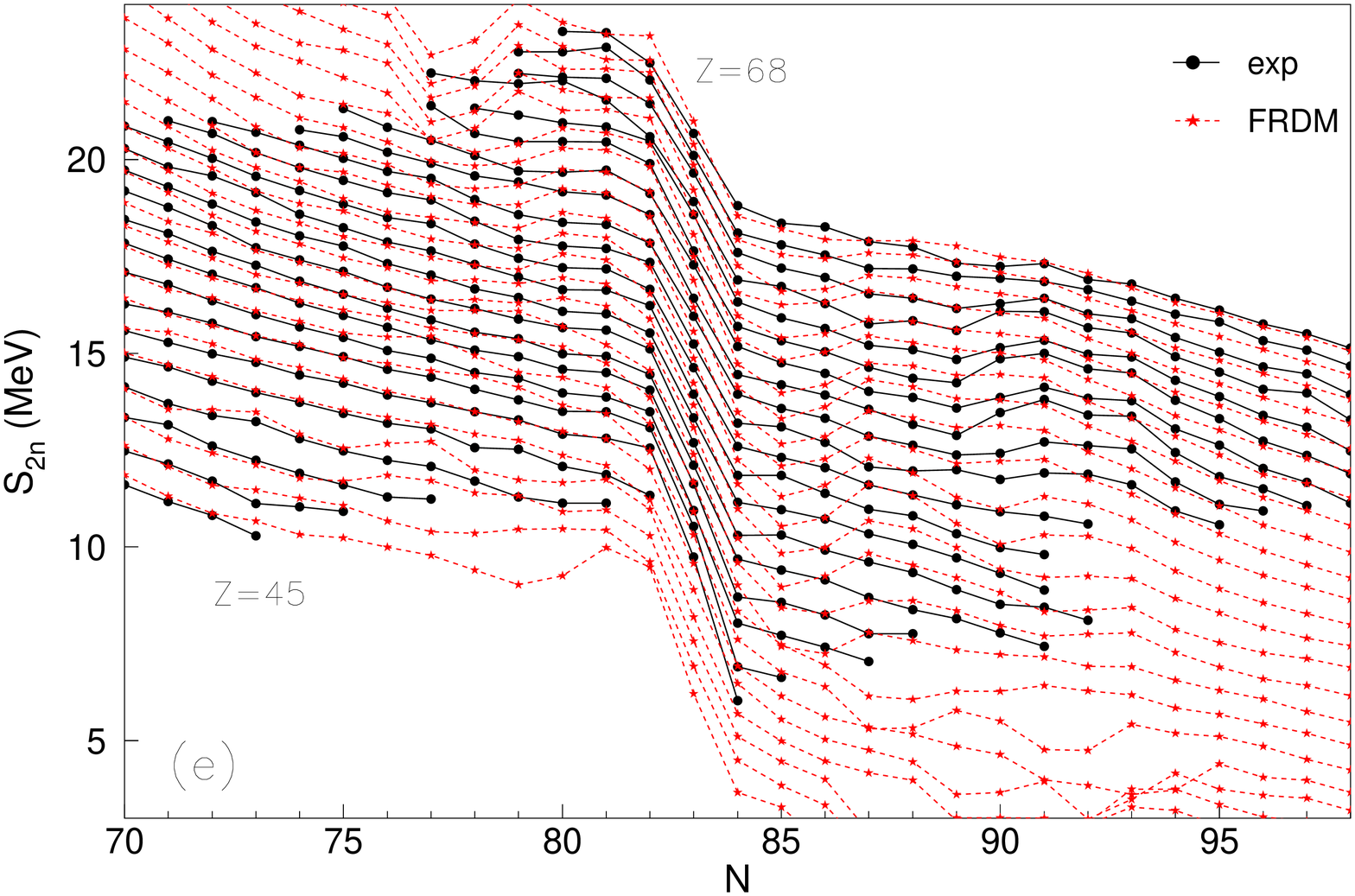}
\includegraphics[angle=0, width=0.35\columnwidth]{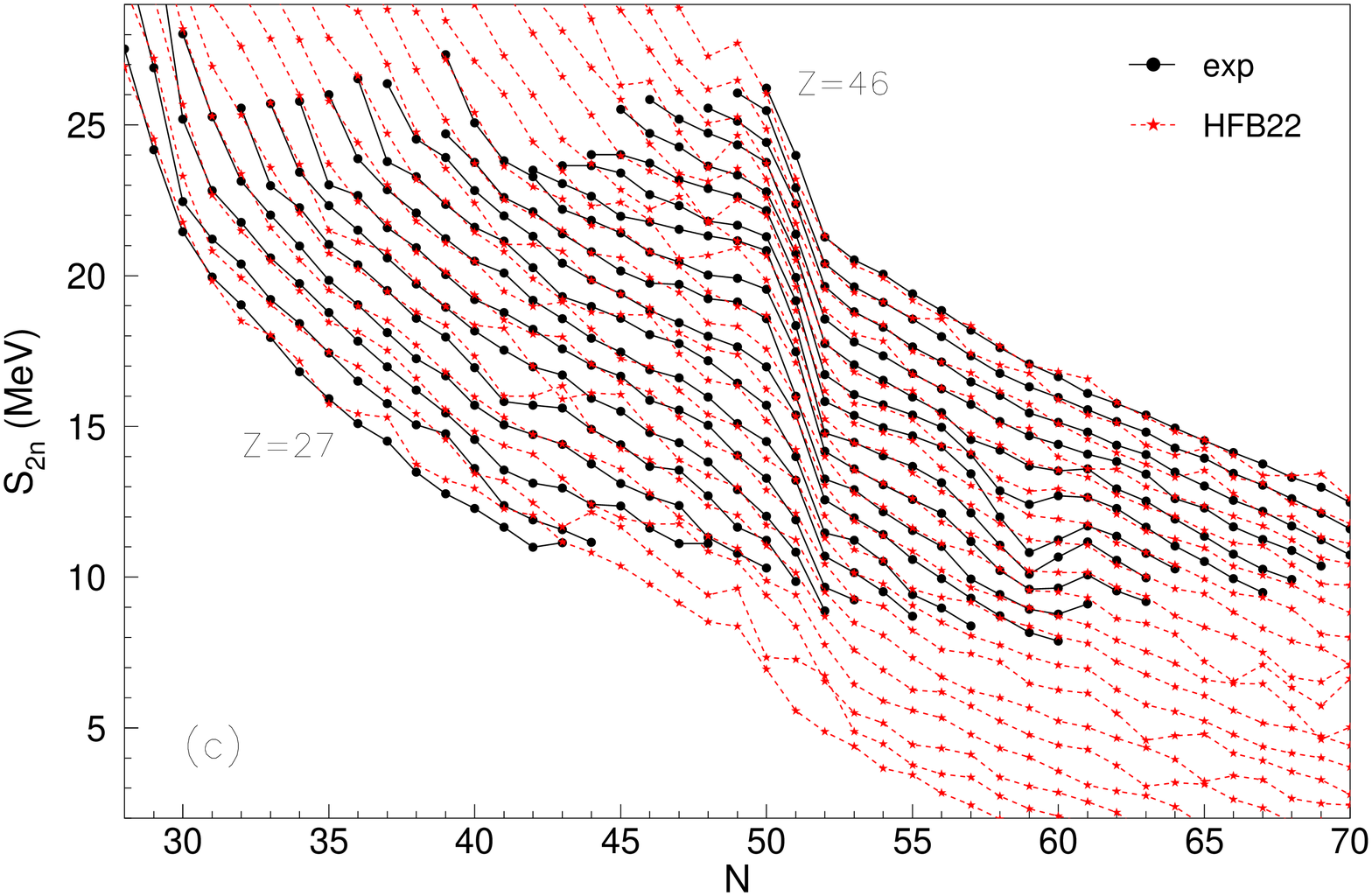}
\includegraphics[angle=0, width=0.35\columnwidth]{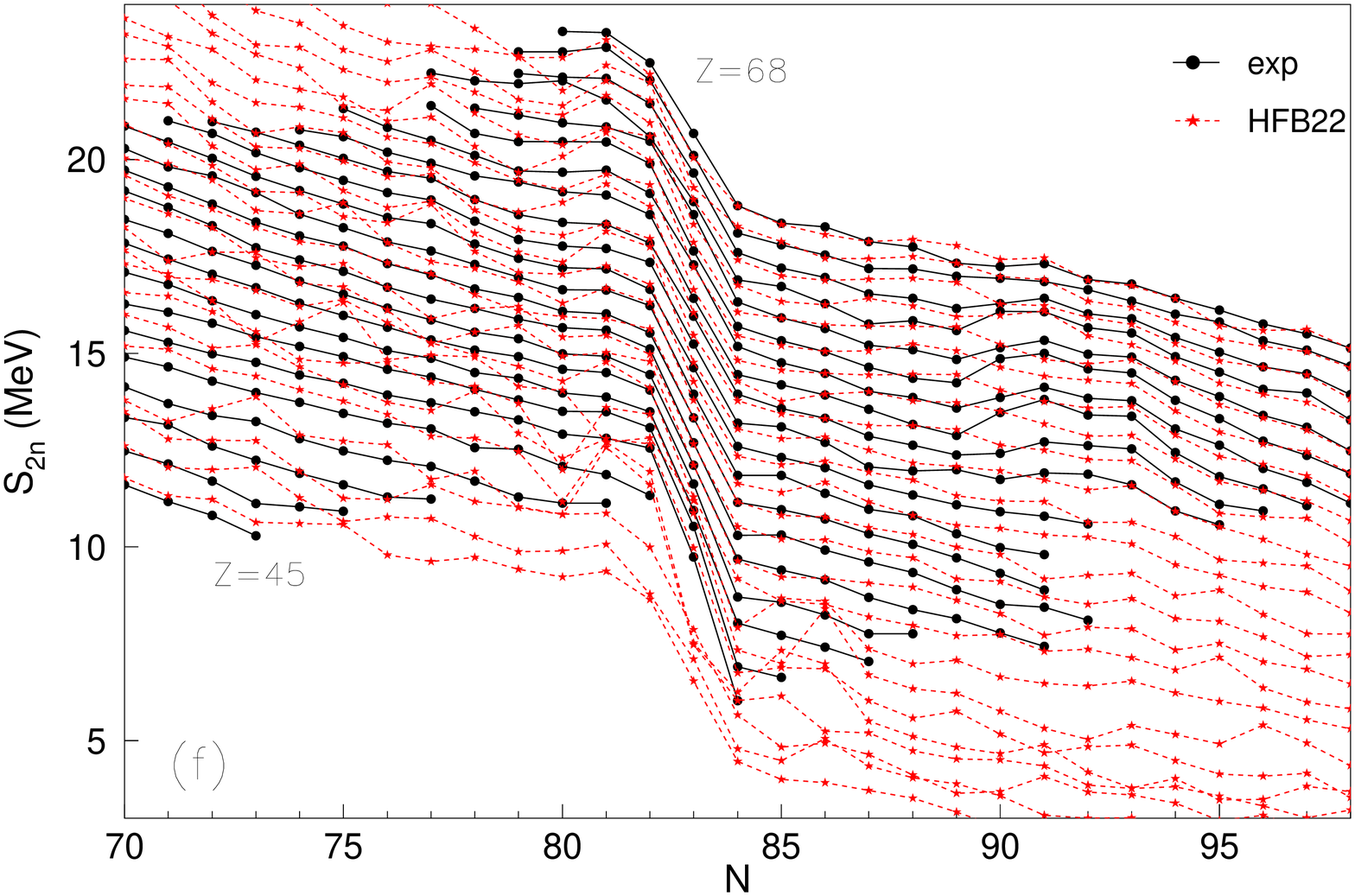}
\end{center}
\caption{(Color online) Two neutron separation energy over selected isotopic chains 
as predicted by different mass models (DZ10 \cite{DZ10}, FRDM \cite{Moller81,Moller93} 
and HFB22 \cite{BsK22}) (solid stars), 
and in comparison to experimental (solid circles) 
from Ref. \cite{Audi}.
} 
\label{fig:mass}
\end{figure}

Since we are here mainly interested in the region around the 
$N=50$ and $N=82$, and in order to get an idea of the performance of
the different mass models, Fig.~\ref{fig:mass} shows the evolution
of the two-neutron separation energy for different elements around the
magic neutron numbers $N=50$ and $N=82$ as predicted by the 10
parameter model of Duflo and Zuker\cite{DZ10} and two other extremely
successful and popular nuclear mass models, the Bruxelles functional
BSK22 \cite{BsK22} and the Finite Range Droplet Model (FRDM) by Moller
and Nix \cite{Moller81,Moller93}, in comparison to experimental data from
Ref.~\cite{Audi}. 
The overall dispersion of the models on the binding energy over the global
mass table is 561 KeV for DZ10, 580 KeV for BSK22, and
656 KeV for FRDM \cite{BsK22,Wang13} 
\footnote{These numbers are obtained on the more restricted AME mass table from 2003.}.

We can observe that all models appear approximately equivalent in the
level of reproduction of experimental data. Their extrapolation in the
neutron rich region however increasingly deviate from each other approaching the
dripline. In particular, a clear and strong slope change is visible at
$N=50$ and $N=82$ for all elements including the lightest ones for the
phenomenological models DZ10 and FRDM. 
This indicates the presence
of a very strong magic number, unmodified by the increasing neutron
richness. Conversely the more microscopic BSK22 shows a more irregular
behavior, and a certain quenching of magicity going towards the neutron
dripline~\cite{Dobaczewski94,Dobaczewski95}, especially for $N=50$. 
This shell quenching appears to
be closely related to the treatment of the pairing interaction. 
A proper inclusion of pairing correlations is certainly a very delicate issue, 
but  one can at least expect that it should be better treated in a self-consistent 
Hartree-Fock-Bogoliubov calculation such as BSK22 than in a phenomenological mass model.  
These differences show the difficulty of extrapolation of nuclear masses far
from stability.

\newpage 

\onecolumngrid

\twocolumngrid

\newpage

\section{Results}
\label{sec:sec3}
  
\subsection{Thermodynamic conditions during core collapse}
\label{sec:thermo}

\begin{figure}
\begin{center}
\includegraphics[angle=0, width=0.98\columnwidth]{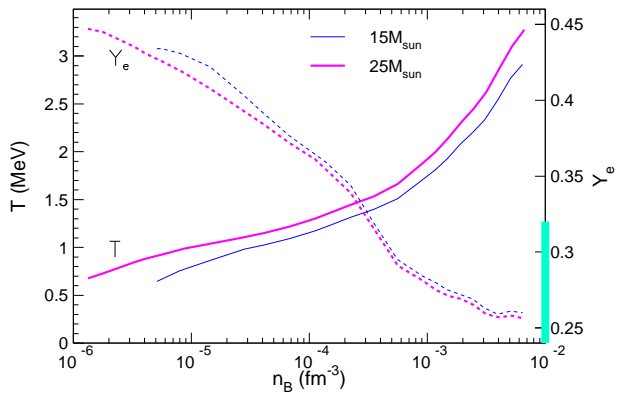}
\end{center}
\caption{(Color online) Thermodynamic conditions ($T, n_B, Y_e$ )
  reached by the central element in the core-collapse of two
  progenitor stars with zero age main sequence masses $15 M_{\odot}$
  and $25 M_{\odot}$ as reported in
  Ref.\cite{Juodagalvis_2010}. 
  Simulations are from Refs.~\cite{Heger01}. 
}
\label{fig:thermocond}
\end{figure}

Exact values of temperature, baryon number density and proton
fraction reached during core collapse depend on many ingredients and
among others the chosen EoS and weak interaction rates. Thus,
without performing a simulation, the results cannot be completely
consistent. We think, however, that it is sufficient to take typical
values for the purpose of the present paper, where we aim to
illustrate the possible impact of a modification of nuclear masses
far from stability on the electron capture rates. A detailed
simulation is left for future work.
 
To obtain such typical values, we consider here two different core
collapse trajectories from Refs.~\cite{Heger01}, as
reported by Juodagalvis et al. in Ref. \cite{Juodagalvis_2010}.  
They correspond to the pre-bounce evolution for the central element of the star at an
enclosed mass of 0.05 solar masses, in the case of a $15 M_{\odot}$
and a $25 M_{\odot}$ progenitor. These simulations use rates 
from LMSH tables~\cite{Langanke00,LMP} and the fully
general relativistic, spherically symmetric AGILE-
BOLTZTRAN code with detailed neutrino radiation hydrodynamics. 
For further details on the simulations, see Refs.~\cite{Heger01}.  A more complete
study would certainly require a complete and consistent simulation. We
expect, however, that the qualitative findings do not depend strongly
on the quantitative details. In addition, it was observed
~\cite{Liebendorfer05} that the electron fraction profiles are well
correlated with the density during the collapse phase, meaning that
matter, independently of the exact position inside the star, will
follow similar trajectories and that our example conditions taken from
the central element are valid more generally.

The thermodynamic conditions during the evolution in terms of
temperature, baryon number density and proton fraction are shown in
Fig.~\ref{fig:thermocond}. The vertical bar indicates the region in
$Y_p = Y_e$, where the experimental information on nuclear masses starts to
be incomplete according the nuclear abundances obtained within our
NSE model employing experimental masses~\cite{Audi} complemented
with the DZ10 mass table~\cite{DZ10}.  For both progenitors, this
happens at densities higher than $n_B\approx 4 \times 10^{-4}$ fm$^{-3}$,
and temperatures above $T \approx 1.4 $ MeV.  We expect that the mass
model has a considerable influence at this stage of the collapse.

\subsection{Chemical composition and capture rates}
\label{sec:chemical}
\begin{figure}
\begin{center}
\includegraphics[angle=0, width=0.98\columnwidth]{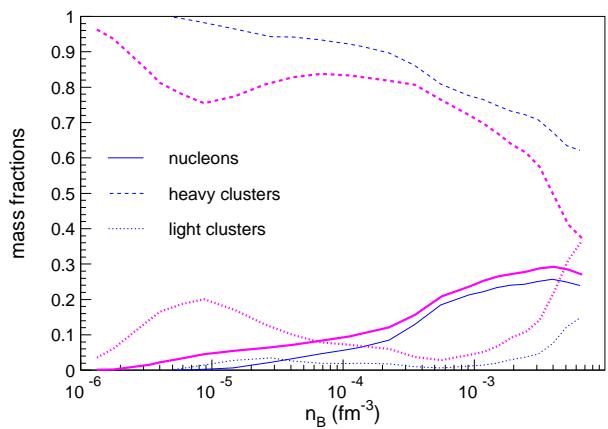}
\end{center}
\caption{(Color online) Mass fractions of free nucleons, light ($2
  \leq A <20 $) and heavy  ($A \geq 20$) clusters obtained for the
  central element of the core-collapse with a $15 M_{\odot}$  
   (thin blue lines) and a $25 M_{\odot}$ (thick magenta lines) progenitor, see Ref.
  \cite{Juodagalvis_2010}. The temporal evolution is labeled via the
  baryon number density as in Fig.~\ref{fig:thermocond}.  The color
  coding is the same as in Fig. \ref{fig:thermocond}.  }
\label{fig:massfrac}
\end{figure}

At any time during collapse, and for both collapse trajectories, the
cluster mass distribution within the NSE model, using experimental
masses~\cite{Audi} complemented with the DZ10 mass table~\cite{DZ10}
for the nuclei for which experimental mass measurements are not
available, has a multi-peaked structure.  A first light cluster
component exponentially decays up to a minimum around $A\approx 20$;
then the abundances increase and show one or several peaks in regions
of the mass table which evolve according to the collapse time.  To
have a global glance of these distributions, Fig.~\ref{fig:massfrac}
displays the mass fraction corresponding to nucleons, light clusters
($2 \leq A < 20$) and heavier clusters ($A \geq 20$). The
thermodynamic conditions are labeled by the baryon number density.
Other mass models lead to very similar results, and we will keep DZ10
as our fiducial mass model in the subsequent calculations.  We can see
that the higher progenitor mass leads to a higher mass fraction for
the lighter elements (nucleons and light clusters), while heavy nuclei
are largely dominant at all times with the less massive progenitor at
this stage of the collapse.  The reason is the systematically higher
temperature in the central element of the more massive star (see
Fig.~\ref{fig:thermocond}).  One can observe, too, that the behavior
of the clusters abundances is non-monotonic for the  {conditions
  obtained with the }higher mass progenitor.    This
is due to the interplay between the increasing temperature during
collapse, which favors light clusters with respect to heavy ones, and
the increasing density leading to an opposite effect.
 
\begin{figure}
\begin{center}
\includegraphics[angle=0, width=0.75\columnwidth]{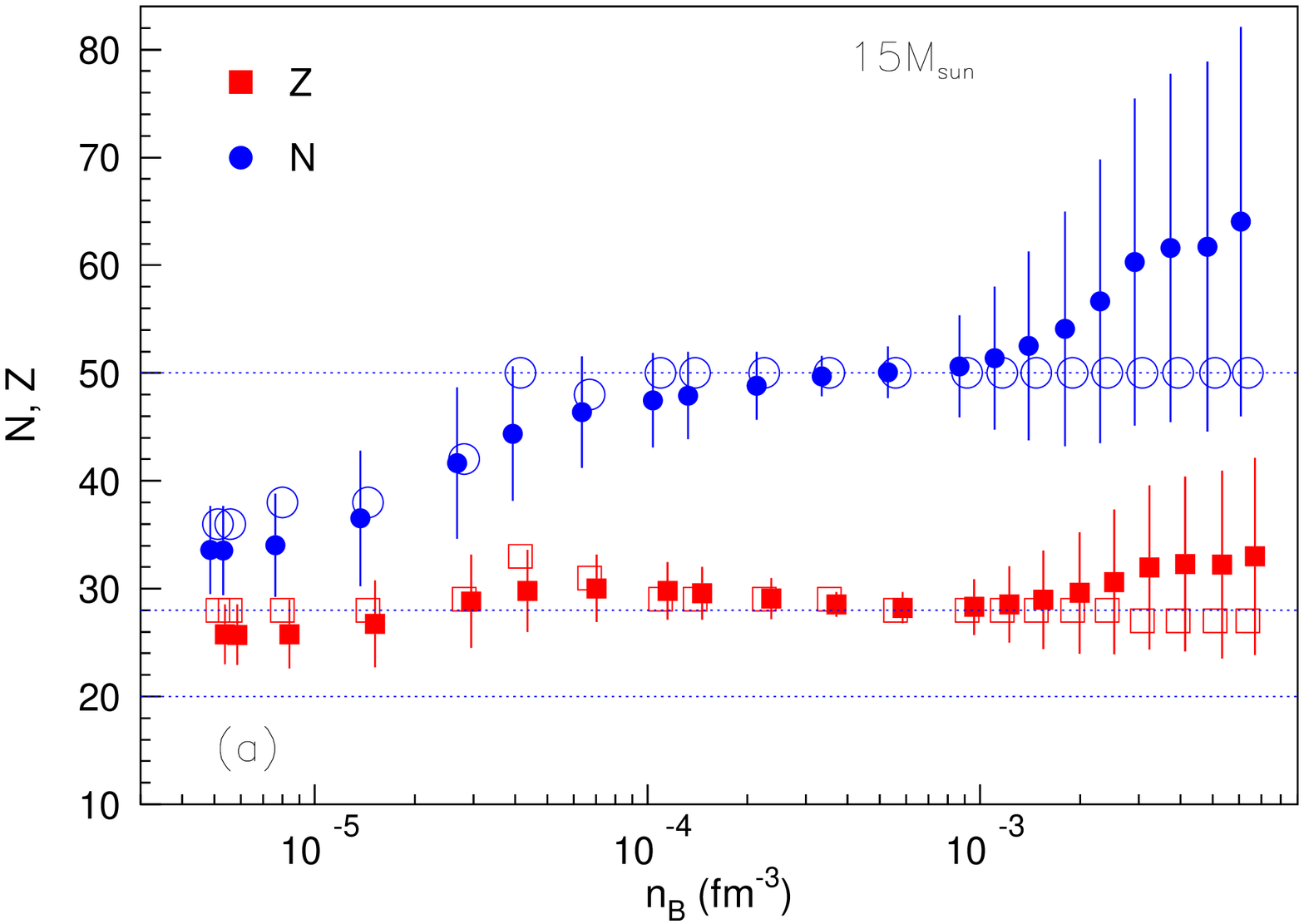}
\includegraphics[angle=0, width=0.75\columnwidth]{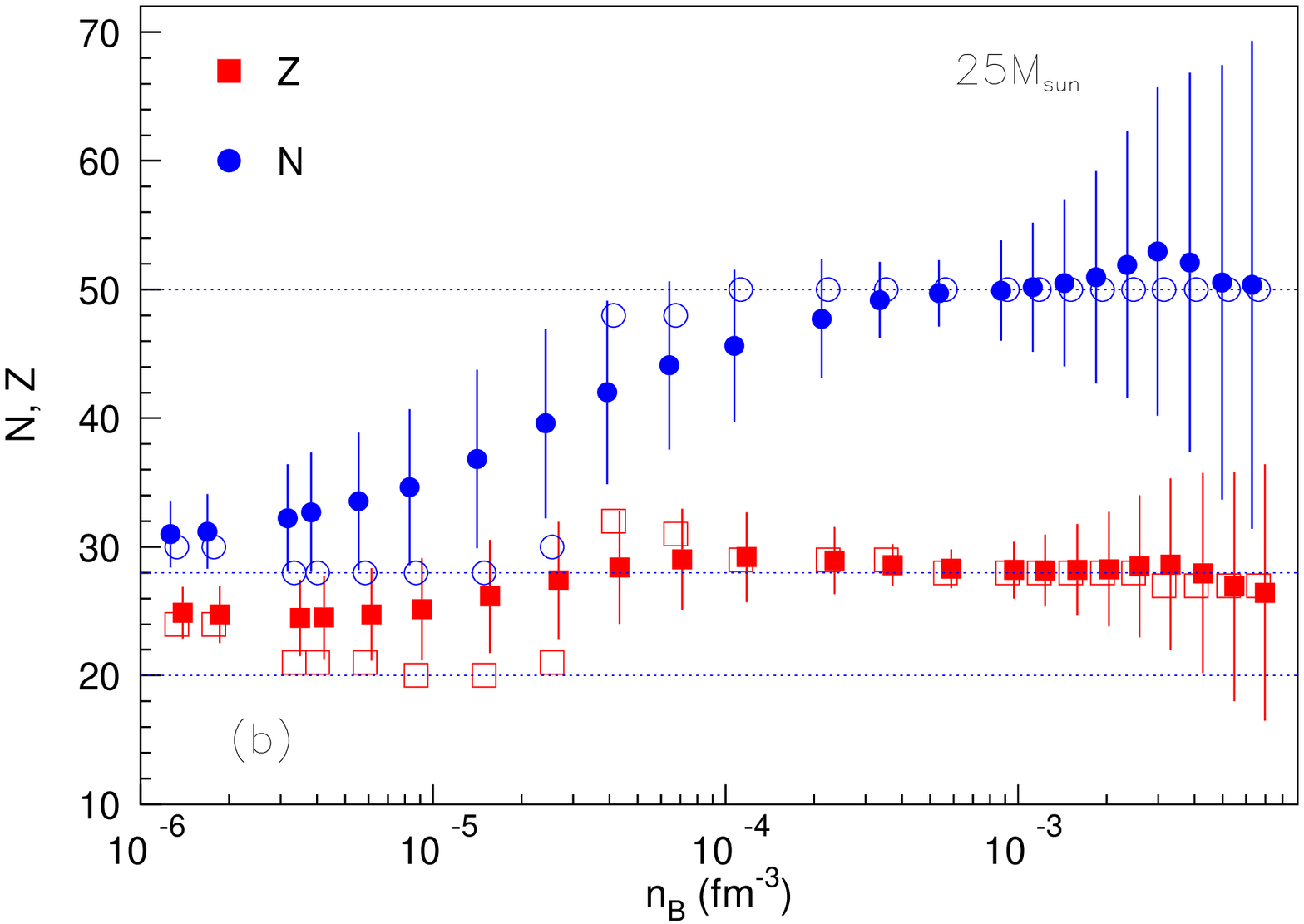}
\end{center}
\caption{(Color online) Average (solid symbols) and most probable
  (open symbols) proton and neutron numbers of heavy ($A \geq 20$)
  nuclei produced in the central element of the core-collapsing $15
  M_{\odot}$ and $25 M_{\odot}$
  progenitors~\cite{Juodagalvis_2010,Hix03,Langanke03} as a function
  of baryonic density.  The vertical bars correspond to the standard
  deviation of the distribution.  For better readability, $N$ and $Z$
  data have been slightly displaced in density. }
\label{fig:azn}
\end{figure}

The composition of heavy ($A \geq 20$) nuclei is further explored in
Fig.~\ref{fig:azn}, which shows for the two trajectories the average
and most probable neutron and proton numbers of these nuclei.  We can
see that the two trajectories produce qualitatively similar patterns.
More precisely, by increasing the density larger clusters are
produced.  Increased densities imply, along the considered
trajectories, increased neutron enrichment of stellar matter which, in
turn, leads to increased neutron enrichment of nuclear clusters.
The standard deviation of the neutron and proton numbers distribution
 (signaled by the
vertical bars) are never negligible meaning that treating matter
composition within the Single Nucleus Approximation (SNA) would have
produced erroneous results, as already acknowledged by the pioneering
works \cite{Aufderheide94}.  In both cases and over most of the
explored density range average $N$ and $Z$ numbers differ from the
most probable ones.   {One reason is }that the
distributions are not only broad but multi-peaked, too. Indeed, very
often the most probable nuclei are magic in $N$.  Magicity in $Z$ is
less frequent because of the smaller total number of protons.  The
reduced number of protons with respect to neutrons explains also the
systematically narrower distributions in $Z$.  {Finally, the
  increasing width of the distributions in neutron number with
  increasing density arise due to the competition of a second magic
  number, in the present case $N = 82$.}

The stability of the fragment production pattern is in agreement with the findings 
by Ref.~\cite{Sullivan15}, who have obtained a very similar multiplicity distribution
pattern for a bunch of different trajectories with different
progenitor models and EoS. 
We therefore conclude that the present simulation is reasonably representative 
of the generic evolution of the central element. 



\begin{figure}
\begin{center}
\includegraphics[angle=0, width=0.98\columnwidth]{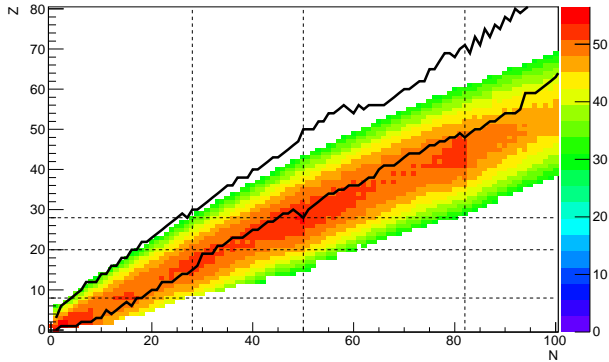}
\end{center}
\caption{(Color online) Nuclear abundances (arbitrary units) corresponding to $n_B=1.18
  \times 10^{-3}$ fm$^{-3}$, $T=2$ MeV, $Y_e=0.275$.  
  The solid lines mark the boundaries of experimental mass measurements.  
  The dotted lines mark magic numbers.}
\label{fig:YNZ_T=2}
\end{figure}

To better understand the origin of this behavior,
Fig.\ref{fig:YNZ_T=2} displays the isotopic abundances corresponding
to $n_B=1.18 \times 10^{-3}$ fm$^{-3}$, $T=2$ MeV, $Y_e=0.275$,
{\it i.e.} to the time where the dispersion of the distribution of
Fig.~\ref{fig:azn} starts to  {become non-negligible.}
  The distribution is centered around the
$N=50$ neutron magic number, and the important width is due to the
emergence of a second peak around $N=82$. This finding is in agreement
with the simulations in Ref.~\cite{Sullivan15} and a similar effect
was observed for the neutron star crust in
Refs.~\cite{Wolf13,Kreim13}. In particular, in Ref.~\cite{Sullivan15}
it was shown that the overall variation of the electron fraction
during the collapse is most sensitive to the electron capture rate on
nuclei in the mass range 74$ \leq A \leq$84, particularly on
$^{78}$Ni, $^{79}$Cu, and $^{79}$Zn close to the $N = 50$ shell
closure.  Our findings confirm these results. The two lines in
Fig.\ref{fig:YNZ_T=2} show the borders of the region where mass
measurements exist, though with a variable degree of precision. We can
see that the most abundant nuclei lay just outside this border.  This
means that their abundance, and therefore their preeminent role in the
electron capture mechanism, relies on the extrapolation of the $N=50$
and $N=82$ shell closure far from stability, in a neutron rich region
where mass measurements do not exist and spectroscopic informations
are scarce and incomplete.
This means that small modifications in the nuclear binding
energies of nuclei with masses not yet measured, but measurable in a
near future, can change star matter composition and, consequently, all
astrophysical quantities depending on it, notably weak rates. This
  point will be explored in more detail in the following subsection.

\begin{figure}
\begin{center}
\includegraphics[angle=0, width=0.8\columnwidth]{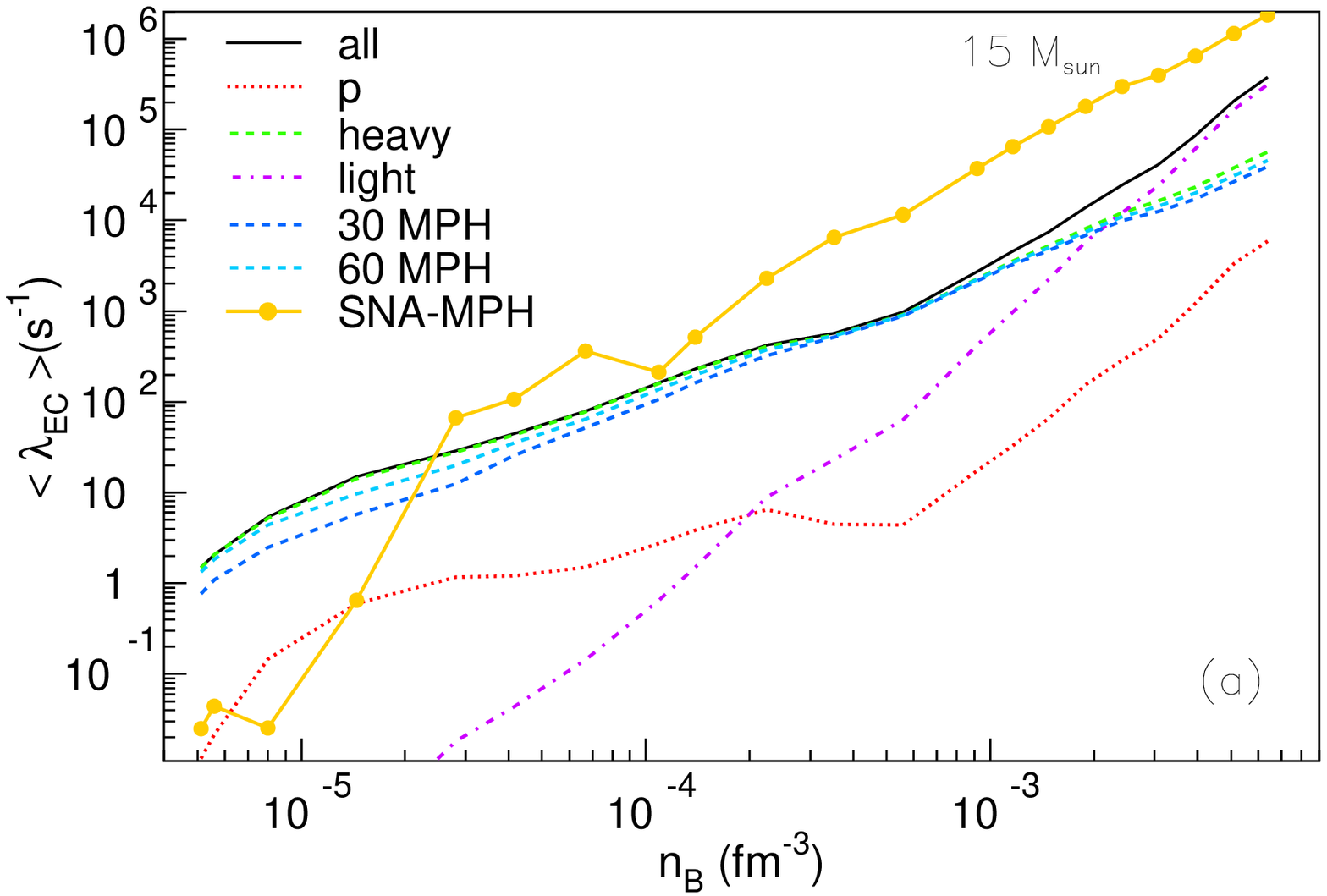}
\includegraphics[angle=0, width=0.8\columnwidth]{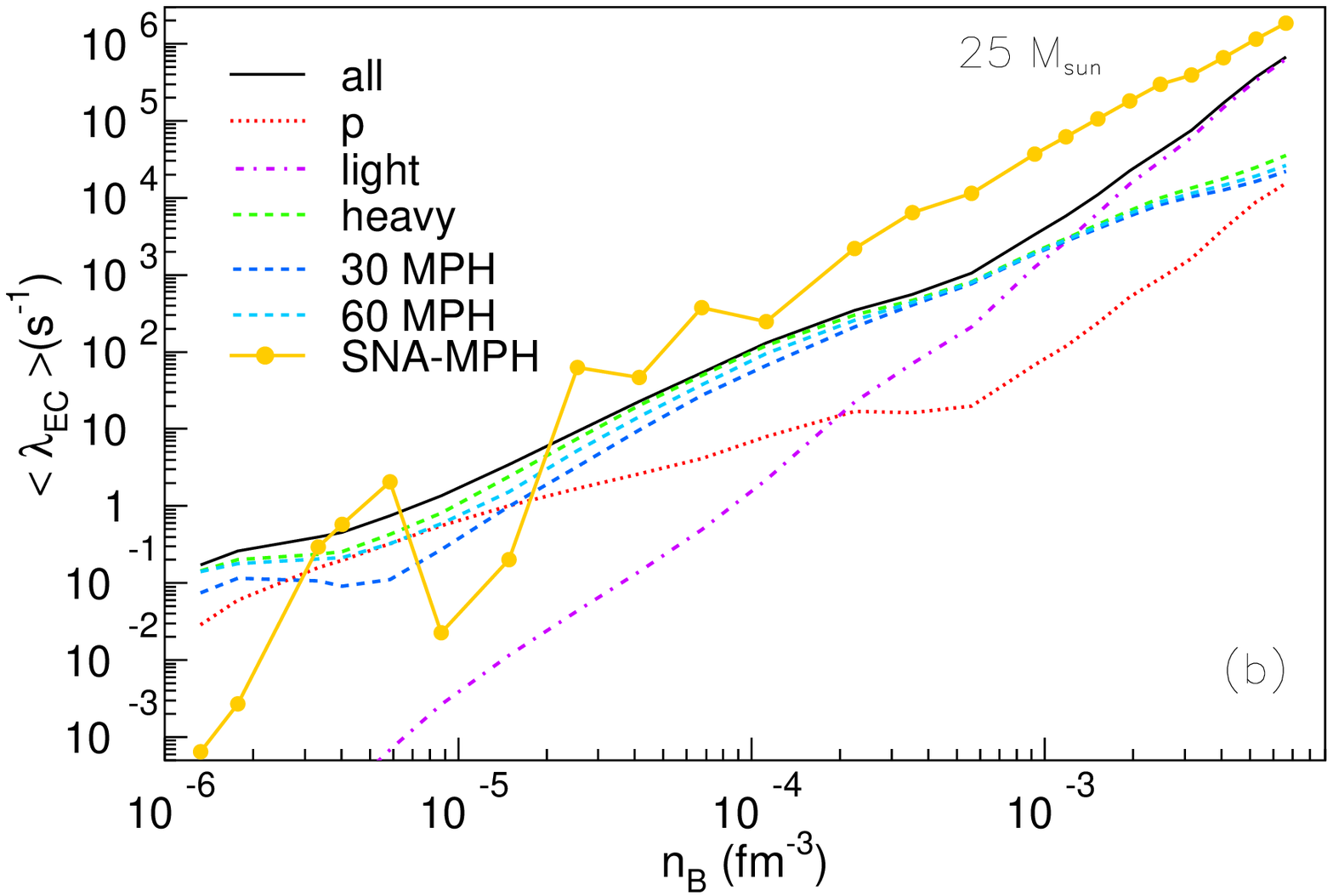}
\end{center}
\caption{(Color online) Average electron capture (EC) rates on
  protons, light ($2 \leq A < 20$) nuclei, heavy ($A \geq 20$) nuclei
  and the 30 (dashed blue lines) and 60 (dashed cyan lines), respectively, most probable heavy nuclei.
  SNA-like EC rates (see text) are also plotted (SNA-MPH).  
  The considered thermodynamical trajectories correspond to the central
  mass element of the $15M_{\odot}$ and $25M_{\odot}$-progenitors\cite{Juodagalvis_2010}. 
}
\label{fig:ec}  
\end{figure}

For this fiducial model, we now show the NSE averaged electron capture
(EC) rates on the different species ${\cal C}$, 
$\langle \lambda_{EC}^{\cal C} \rangle=\sum_{(N,Z) \in {\cal C}} n^{(N,Z)} \lambda_{EC}^{(N,Z)}/\sum_{N,Z} n^{(N,Z)}$.  
The two above mentioned
core-collapse trajectories will be considered and L03 formulae (see
Sec.~\ref{sec:individual}) will be used.  Let us first concentrate on
the capture rates on heavy nuclei.  As it is well known \cite{Aufderheide94}, 
a very huge number of different nuclear species contribute to the total rate
in all thermodynamics conditions. This can be appreciated by
limiting the rate calculation to  the $N$ most probable nuclei, 
$\langle \lambda_{EC}^{N}\rangle=\sum_{i=1}^N n_{i} \lambda_{EC}^i/\sum_{N,Z} n^{(N,Z)}$,
where $n_i$ is the abundance of the $i-th$ most probable
cluster. Comparing the result obtained with $N=30$ and $N=60$,
with the one corresponding to the whole distribution, one can see how important it
is to properly account for the complete distribution of nuclear
species. In the case of both progenitors the 60 most probable nuclei
never exhaust the average EC rate on heavy nuclei
and, for instance, at $n_B=1.4 \times 10^{-5}$ fm$^{-3}$ they account for only 60\% 
of $\langle \lambda_{EC}^{heavy} \rangle$.

The inadequacy of the single nucleus approximation was recently
stressed in Ref.\cite{Hempel12}. In that work it was shown that
sizable differences in the collapse dynamics are obtained if the NSE
model is replaced with a more conventional model \cite{STOS}
considering a single representative Wigner-Seitz cell for each
thermodynamic condition, even if the same TM1 energy functional was
employed in both models. However, in that work the individual rates
were replaced by a single rate on the most probable cluster, using the
simplified Bruenn
parametrization~\cite{Bruenn85}. Our results show that the complete
nuclear distribution should also be used in the calculation of the
rates.  If we replace the folding of the individual rates with the EC
rate of the most probable nucleus weighted by the baryon number fraction bound
in clusters,  {$\langle \lambda_{EC}^{SNA}\rangle=\left(n_{cl}/n_B \right)
  \lambda_{EC}^{MP}$}, the result (lines with points) is seen to very
badly reproduce the complete folding result.  As one may see in
Fig. \ref{fig:ec}, at low densities $\langle\lambda_{EC}^{SNA}\rangle$ 
generally underestimates the NSE averaged EC rate and, 
because of structure effects related to the low temperatures at these densities,
manifests a huge scattering. On the contrary, at higher densities and
temperatures ($n_B>3 \times 10^{-5}$ fm$^{-3}$),  
$\langle\lambda_{EC}^{SNA}\rangle$ largely overestimates the NSE-averaged EC rate.
Obviously, different values are expected if instead of the L03 approximation,
other prescriptions are employed for the individual EC rates, see
the discussion in Ref.~\cite{Sullivan15}, too. However, the general
trend induced by the different averaging procedures should remain
the same.

Concerning the EC rates on light nuclei, they are negligible at the beginning
of the collapse but increase strongly, mainly due to the increasing
temperature, and become dominant in the latest stage. This underlines
the importance of including microscopic calculation of EC rates
for light nuclei, too \cite{Fuller10}.

The total rates are given by the sum of the different contributions,
including electron capture on free protons (dotted lines) which
however plays a minor role at all times.  Comparing
Figs. \ref{fig:thermocond}, \ref{fig:azn} and \ref{fig:ec}, we note
that for thermodynamic conditions where very neutron rich nuclei start
to dominate, with masses which are not experimentally known, they
still represent the major electron capture source.  This qualitative
picture is independent of the progenitor mass.  From a more
quantitative point of view, the relative importance of heavy clusters
is higher for the lower mass progenitor, essentially because of the
lower temperatures reached during the collapse.

\subsection{Evolution of magicity far from stability}

\label{sec:evolution}
It is well known in the recent nuclear structure literature that even
major shell closures can be quenched far from stability, see
e.g.~\cite{Sorlin08,Chaudhuri13,Sorlin12}. A very clear evidence exists for the $N=20$
magic number \cite{Sorlin08,Chaudhuri13}, which corresponds to a huge
gap for decreasing proton number up to $^{34}$Si, and suddenly
disappears in the next even-even isotope $^{32}$Mg. This is partly due
to the modification of single-particle energies far from stability,
with the consequence that new magic numbers can appear corresponding
to strongly deformed configurations. However the main reason of the
modification of magicity relies on effects which go beyond the naive
single particle shell model. First, the effect of the proton-neutron
residual interaction in nuclei with strong asymmetry is decreased;
moreover, correlations play an increasing role which makes the very
concept of shell closure less relevant. As a consequence, the main
effect is a quenching of the shell gap, even if secondary new gaps can
appear in very localized regions of the nuclear chart.

The possible modification of the $N=50$ and $N=82$ shell gaps far from
stability is the object of intense theoretical and experimental
research in nuclear structure (e.g.~\cite{Sorlin08,Sorlin12}).  Here,
we do not have the ambition to model this phenomenon, but simply
analyze the modifications in the matter composition and associated
electron capture rates, which would be induced by the expected shell
quenching.

A similar idea was proposed in Ref.~\cite{Pearson96}, who showed that a
modified mass formula built to incorporate the possibility of shell
quenching, has a striking impact on canonical calculations of the
r-process. We follow a similar strategy as in Ref.~\cite{Pearson96} and
introduce a modified expression for the binding energy in the following form:

\begin{eqnarray}
B^m(A,Z)&=& B^{exp}(A,Z), ~~ Z_i^{exp}(A) \leq Z \leq Z_s^{exp}(A), \nonumber \\
&=& B^{LD}(A,Z)+f(Z_i^{exp}(A)-Z, \Delta Z, \alpha) \nonumber \\
&\times& \left(B^{DZ}(A,Z)-B^{LD}(A,Z)\right), \nonumber \\
&~&  Z_i^{DZ} (A) \leq Z < Z_i^{exp}(A) \nonumber \\
&=& B^{LD}(A,Z)+f(Z-Z_s^{exp}(A), \Delta Z, \alpha) \nonumber \\
& \times& \left(B^{DZ}(A,Z)-B^{LD}(A,Z)\right), \nonumber \\
&~&  Z_s^{exp} (A) < Z \leq Z_s^{DZ}(A), \label{eq:bmod} 
\end{eqnarray}
 
where $B^{exp}(A,Z)$ and $B^{DZ}(A,Z)$ stand for the experimental binding
energy~\cite{Audi} and, respectively, predictions of DZ10 mass
model~\cite{DZ10}.  $Z_i^{exp}(A)$ (resp., $Z_i^{DZ}(A)$) and $Z_s^{exp}(A)$
(resp. $Z_s^{DZ}(A)$) correspond to the most neutron-rich and, respectively,
most neutron-poor nucleus with $A$ nucleons for which experimental
masses (resp., predictions of DZ10) exist.
  $B^{LD}(A,Z)$ is a simple
liquid-drop binding energy calculated according to

\begin{eqnarray}
  B^{LD}(A,Z)&=& a_v A-a_s A^{2/3}-a_{vi} 4 I(I+1)/A \nonumber \\
  &+& a_{si} 4 I(I+1)/A^{4/3} - a_c Z(Z-1)/A^{1/3} \nonumber \\
    &+& V_p(A,Z),
\end{eqnarray}
with 
$I=|A-2Z|/2$,
$a_v$=15.62 MeV, $a_s$=17.8 MeV, $a_{vi}$=29 MeV, $a_{si}$=38.5 MeV, 
$a_c$=0.7 MeV and $V_p=\pm 12/\sqrt{A}$ MeV for even-even (+) and, 
respectively, odd-odd nuclei (-).

Finally we introduce a smearing function depending on the parameters $\Delta Z$
and $\alpha<0$
which determine how sudden the shell quenching is supposed to be:

\begin{eqnarray}
f(x, \Delta Z, \alpha)=\exp\left[\alpha x/\Delta Z \right].
\end{eqnarray}

Small values of $\Delta Z$ correspond to maximum quenching, while in the 
limit $\Delta Z \to \infty$ we recover the DZ10 functional form, 
which predicts preserved magic numbers up to the dripline.
 
\begin{figure}
\begin{center}
\includegraphics[angle=0, width=0.98\columnwidth]{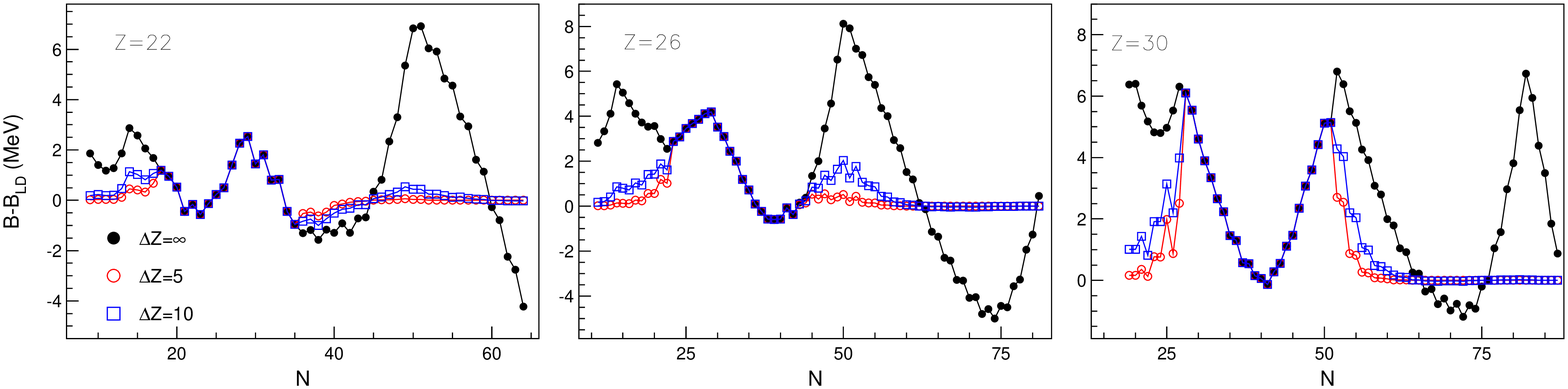}
\includegraphics[angle=0, width=0.98\columnwidth]{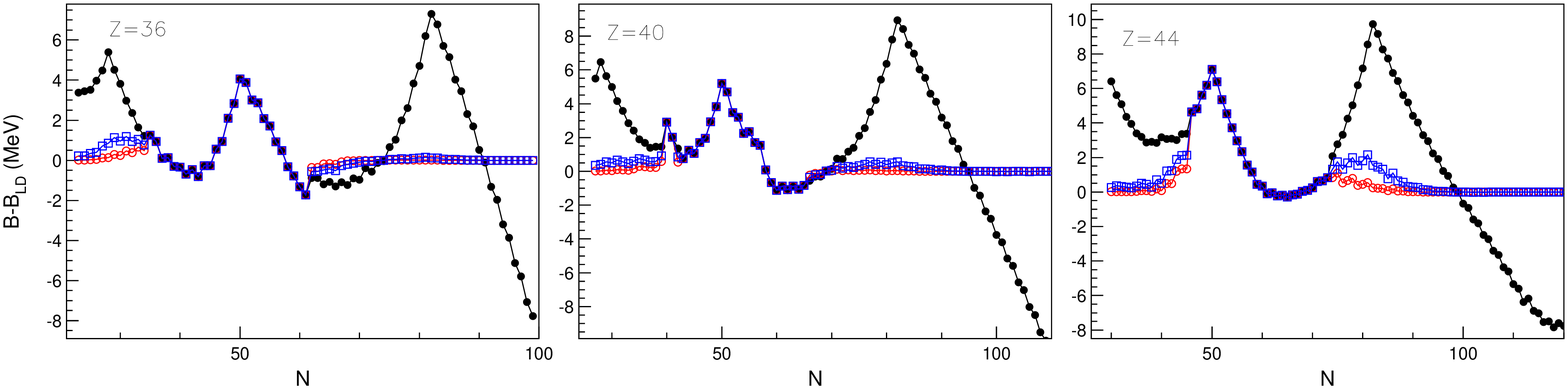}
\end{center}
\caption{(Color online) LDM-shifted binding energy as a function of neutron number
  for different isotopes strongly populated during core
  collapse. DZ10~\cite{DZ10} results (solid black dots) are plotted
  along modified results $B^m$ corresponding to two different scenarios
  of shell quenching (see text for details). }
\label{fig:Bmodif}
\end{figure}

The behavior of the modified functional on the binding energy is illustrated in
Fig.~\ref{fig:Bmodif} for different arbitrary values of the smearing
parameter $\Delta Z=5, 10, \infty$ and $\alpha=\log \left(10^{-2}\right)$.
A selected number of nuclei is displayed as a function of neutron
number. These elements have been chosen since they are strongly
populated in the later phase of the collapse.

As observed before, the DZ10 model shows an extremely pronounced shell
closure at $N=50$ and $N=82$ for all elements including the exotic
ones like $^{72}$Ti and $^{118}$Kr. In the modified expression,
Eq.~(\ref{eq:bmod}), the gap is quenched far from stability, and the
quenching is more or less pronounced depending on the choice of the
parameter $\Delta Z$. 

\begin{figure}
\begin{center}
\includegraphics[angle=0, width=0.98\columnwidth]{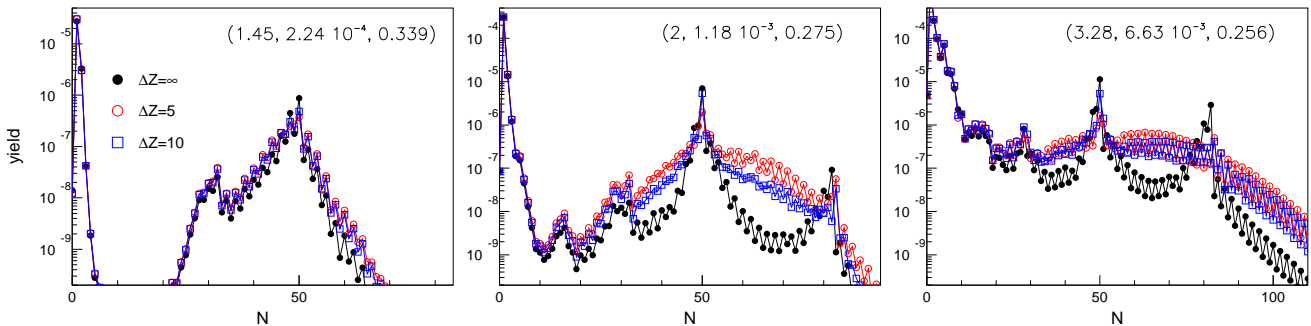}
\end{center}
\caption{(Color online)} 
\label{fig:YN_B} Impact of nuclear binding energy on nuclear abundances: 
the distribution of clusters with a given neutron number is shown.
The considered thermodynamic conditions  {($T$ [MeV], $n_B$ [fm$^{-3}$], $Y_p$)} are mentioned
on each panel, corresponding to three different times in the evolution
of the central element of the collapse with a 25 $M_\odot$ 
progenitor~\cite{Heger01,Juodagalvis_2010}. The same prescriptions for the
binding energies as in Fig.~\ref{fig:Bmodif} are used.
\label{fig:distrimodif}
\end{figure}

The effect of the modification of the mass formula on the distribution
of nuclei is shown in Fig.\ref{fig:distrimodif} for three different
representative conditions during the collapse, taken from the
25$M_\odot$ progenitor~\cite{Juodagalvis_2010,Hix03,Langanke03}. Obviously, the
effect of shell quenching is to reduce the size of the magic peaks and
favor open shell nuclei, thus leading to a wider isotopic
distribution.

\onecolumngrid

\newpage
\twocolumngrid

\subsection{Effect on the electron capture rates}
\label{sec:effect}

\begin{figure}
\begin{center}
\includegraphics[angle=0, width=0.98\columnwidth]{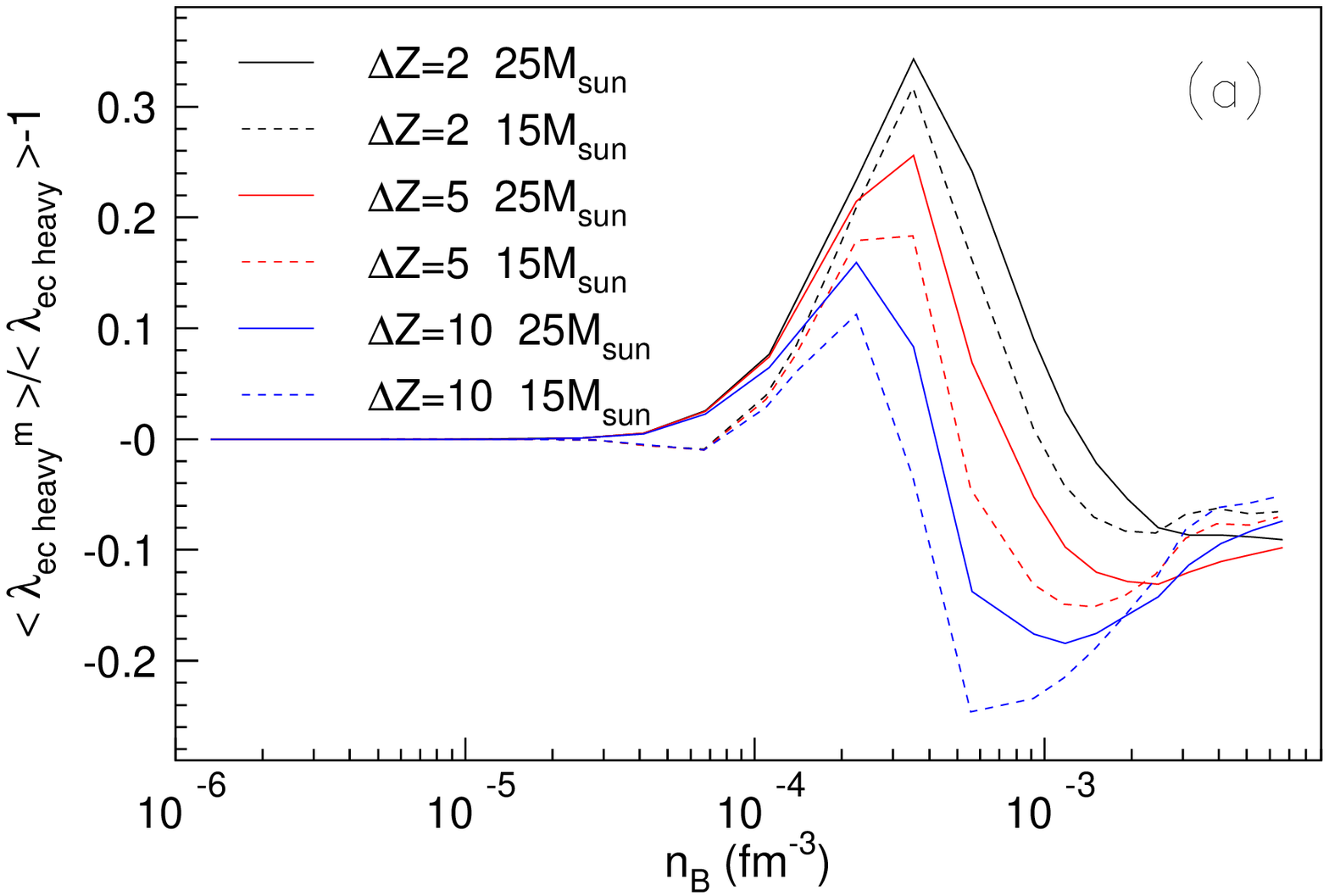}
\includegraphics[angle=0, width=0.98\columnwidth]{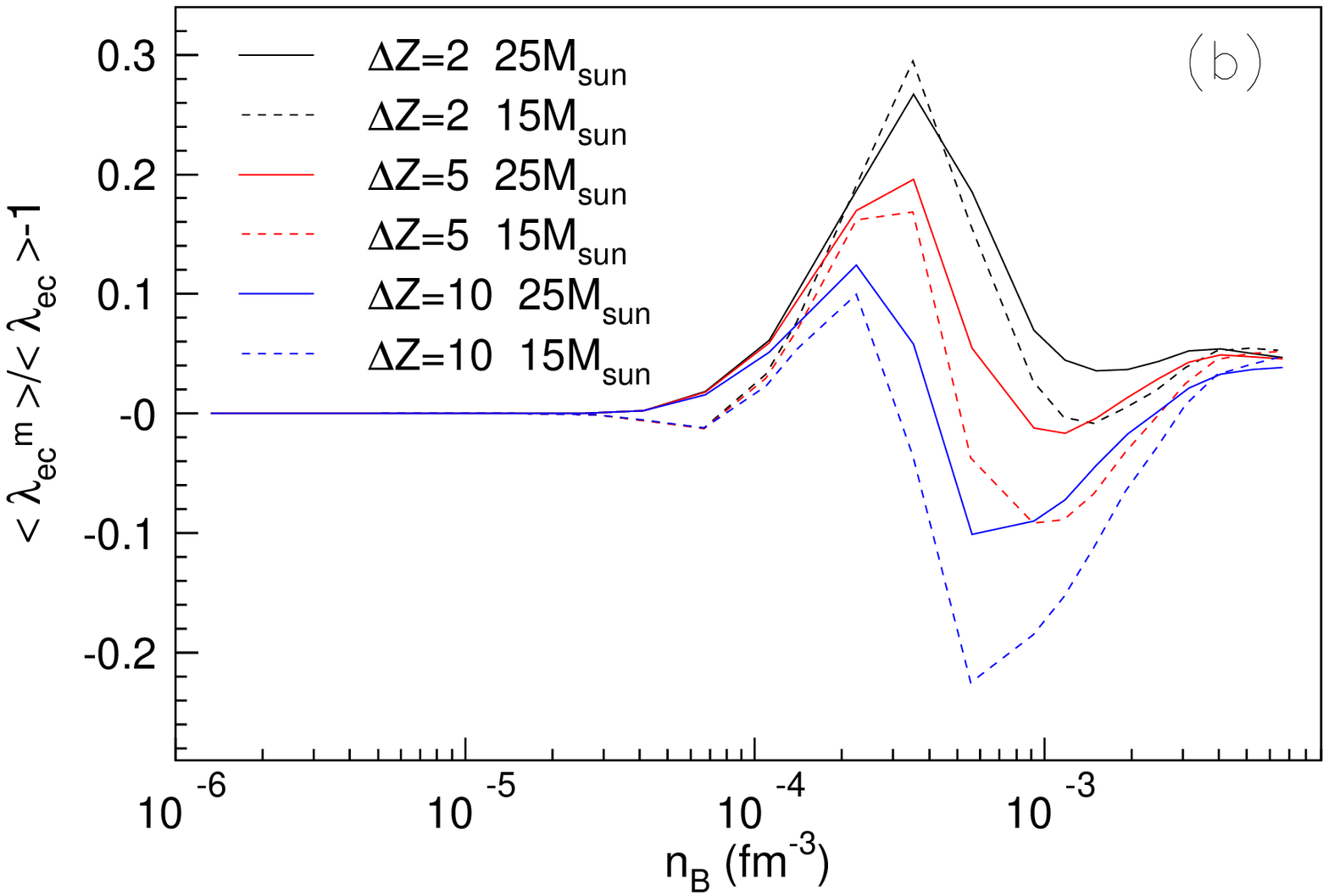}
\end{center}
\caption{(Color online) Ratio between NSE-averaged EC rates using the
  shell-quenched mass functional (see text) and the original
  DZ10~\cite{DZ10} mass model. The thermodynamic conditions are taken
  from the two core collapse trajectories of the central element of a
  15$M_{\odot}$ (dashed lines) and a $25M_{\odot}$ (full lines)
  progenitor~\cite{Heger01,Juodagalvis_2010}. The temporal evolution is
  labeled by the baryon number density as before. The averaged rate is
  calculated only on heavy nuclei ($A \geq 20$) in the upper panel and
  on all nuclei in the lower panel. Different quenching factors
  $\Delta Z$ are considered (see text).  }
\label{fig:ec_modif}  
\end{figure}

Our final result about the impact of shell quenching towards the
drip-line on NSE-averaged electron capture rate is shown in
Fig.~\ref{fig:ec_modif}. 
Several magicity quenching recipes have been considered
$\Delta Z=2, 5, 10$ and $\alpha=\log(10^{-2})$ for each of the two progenitors.
Their predictions are plotted in terms of relative deviations with
respect to the fiducial model.

Panel (a) corresponds to NSE-averaged EC rates on heavy nuclei.
Independently on the stiffness of magicity quenching and progenitor mass, the same
pattern is obtained. Over a certain time after neutron-rich nuclei start to be
copiously populated shell quenching induces an increase of the EC rate.
This is due to a significant abundance increase for non-magic nuclei 
and a moderate abundance decrease of the N=50 magic nuclei.
In this regime  shell quenching on the EC rates on heavy nuclei 
is more pronounced, and lasts for a longer time, in the case of the 25$M_\odot$ progenitor. 
Toward the end of the considered trajectories, the opposite effect is obtained.
Here shell quenching is responsible for a reduction of one order of magnitude 
in the abundance of N=50 and N=82 magic nuclei, not compensated by the 
abundance increase of non-magic nuclei and the corresponding increase in the $Q$-value.
In the two regimes, the amplitude of the effect depends on both  
quenching parameter $\Delta Z$ and progenitor mass and may amount up to
30\%.
Panel (b) shows that, despite the fact that heavy nuclei represent only a fraction
of the whole mass and more isospin-symmetric light nuclei are not affected by shell
quenching, modification of EC rate on heavy nuclei survives
in the inclusive rate. Up to $n_B \lesssim 3 \times 10^{-4}$ fm$^{-3}$, where
$X_{heavy} \gtrsim 0.8$ (see Fig. \ref{fig:massfrac}), 
$\langle \lambda_{EC}^m \rangle /\langle\lambda_{EC} \rangle \approx
\langle \lambda_{EC}^{m~heavy} \rangle /\langle\lambda_{EC}^{heavy} \rangle$.
For the highest considered densities the overall modification of EC rates is opposite to that
seen for heavy clusters. 
More precisely, shell quenching here points toward an increase of EC.
This happens because of a more favorable production of light clusters.

\section{Summary and conclusions}
\label{sec:sec4}

In this paper we have examined the consequence of a possible quenching
of the $N=50$ and $N=82$ shell closures on the EC rates during core
collapse.  As basis of the analysis, we have considered the same
typical thermodynamic conditions as in
Ref.~\cite{Juodagalvis_2010}. They correspond to the pre-bounce
evolution of the central element of the star obtained within a
core-collapse simulation using two different progenitors, a
15$M_\odot$ and a 25$M_\odot$ one, from
Refs.~\cite{Heger01}.  In agreement with
Ref.~\cite{Sullivan15} we find that the properties of very exotic
nuclei around these two shell closures is {a} key microscopic
information to predict the evolution of the electron fraction during
collapse.

We have pointed out that a quenching of these shell closures compared
with the popular DZ10 mass model~\cite{DZ10} considerably affects the
nuclear distribution, and consequently the EC rates during
collapse. Using the standard L03 formalism for rates on individual
nuclei, we have analyzed the modification of NSE average EC rates for
different scenarios of shell quenching. Depending on the progenitor mass and
stiffness of shell quenching modifications of EC rates of up to 30\% have been
obtained. 
We expect that such effects, once consistently included in the time dependent 
evolution of the collapse, have a sizable effect on neutrino
emissivity and on the enclosed mass at bounce.
We should mention that the L03 approximation might be questionable in
this exotic region far away from the stable $fp$-shell nuclei where
these formulae were adjusted. We expect, however, that more precise
rates do not change our qualitative findings. The reason is that
the EC rates are considerably suppressed for nuclei with a
pronounced shell closure due to the energy gap in the single
particle states which makes a transition more difficult. 

In this work the quenching effect is governed by the parameter $\Delta
Z$ whose value is chosen in an arbitrary way.  More sophisticated
calculations in microscopic theories, such as modern interaction
configuration shell model calculations in the relevant nuclear chart
region~\cite{Coraggio15} are currently underway, and will give
essential information on the effective importance of the shell
quenching.  However, the last word will be clearly given by
phenomenology. New precise mass measurements  at the edges of the known
isotopic table could allow for a much better
extrapolation towards the neutron rich region in a very near future.
  
Finally, it is important to stress that the modification of
nuclear structure far from stability is expected to influence not only
the nuclear mass, but also the individual EC
probabilities, which should be consistently calculated together with
the mass model and the energy functional describing the self-interaction
of unbound particles. 
Dedicated QRPA calculations are in progress to
achieve this aim \cite{Fantina2015}.


\section*{Acknowledgments}
This work has been partially funded by the SN2NS project
ANR-10-BLAN-0503 and by NewCompstar, COST Action MP1304.  Ad. R. R
acknowledges partial support from the Romanian National Authority for
Scientific Research under grants PN-II-ID-PCE-2011-3-0092 and PN 09 37
01 05 and kind hospitality from LPC-Caen and LUTH-Meudon.

\end{document}